%% file: RGGED.tex
\documentclass[11pt]{article}
\pdfoutput=1

\usepackage{jheppub}
\usepackage{amssymb,amsfonts,amsmath}
\usepackage{enumerate}
\usepackage{mathrsfs}
\usepackage[section]{placeins}
\usepackage{verbatim}
\usepackage[normalem]{ulem}
\usepackage{enumitem}
\usepackage[utf8]{inputenc}

\usepackage{caption}
\usepackage{subcaption}
\allowdisplaybreaks

\usepackage{soul}

\newcommand{\be}{\begin{equation}}
\newcommand{\ee}{\end{equation}}
\newcommand{\bea}{\begin{eqnarray}}
\newcommand{\eea}{\end{eqnarray}}

\newcommand{\CO}{\mathcal{O}}

\newcommand{\CV}{\mathcal{V}}

\newcommand{\bk}{{\bf k}}

\newcommand{\lr}{\left (}
\newcommand{\rr}{\right )}
\newcommand{\ls}{\left [}
\newcommand{\rs}{\right ]}

\newcommand\qt\tau

\newcommand{\p}{\partial}

\renewcommand{\tilde}[1]{\widetilde{#1}}

\newcommand{\eg}{{\it e.g.,}\ }
\newcommand{\ie}{{\it i.e.,}\ }
\newcommand{\vp}{\varphi}
\newcommand{\del}{\partial}

\makeatletter
\renewcommand{\@seccntformat}[1]{\csname the#1\endcsname.\,\,}
\makeatother
\allowdisplaybreaks

\usepackage[normalem]{ulem}

\let \savenumberline \numberline
\def \numberline#1{\savenumberline{#1.}}

\makeatletter
\def\@fpheader{\relax}
\makeatother

\def\bea{\begin{eqnarray}}
\def\eea{\end{eqnarray}}

\usepackage{xcolor}
\definecolor{orange}{rgb}{1.0, 0.55, 0.0}

\definecolor{GRAY}{rgb}{0.36, 0.54, 0.66}

\newcommand{\MN}{N}

\title{\ \vspace{1.6cm} \\
Renormalization of Galilean Electrodynamics}
\author{Shira Chapman${}^{a}$,
Lorenzo Di Pietro${}^{b,c}$,
Kevin T. Grosvenor${}^{d}$, and
Ziqi Yan${}^{e}$}
\emailAdd{s.chapman@uva.nl}
\emailAdd{ldipietro@units.it}
\emailAdd{\\ \qquad\quad\,\,\,kevinqg1@gmail.com}
\emailAdd{zyan@perimeterinstitute.ca}
\affiliation{
    ${}^a$
        Institute for Theoretical Physics, University of Amsterdam \\
        Science Park 904, Postbus 94485, 1090 GL, Amsterdam, The Netherlands  \medskip\\
    ${}^b$
        Dipartimento di Fisica, Universit\`a di Trieste, Strada Costiera 11,
I-34151 Trieste, Italy
\medskip\\
${}^c$
INFN, Sezione di Trieste, Via Valerio 2, I-34127 Trieste, Italy
\medskip\\
    ${}^d$
        Physikalisches Institut and Würzburg-Dresden Cluster of Excellence ct.qmat,\\
Universität Würzburg, 97074 Würzburg, Germany
\medskip\\
    ${}^e$
        Perimeter Institute for Theoretical Physics \\
        31 Caroline St N, Waterloo, ON N2L 2Y5, Canada}
\abstract{We study the quantum properties of a Galilean-invariant abelian gauge theory coupled to a Schr\"odinger scalar in 2+1 dimensions. At the classical level, the theory with minimal coupling is obtained from a null-reduction of relativistic Maxwell theory coupled to a complex scalar field in 3+1 dimensions and is closely related to the Galilean electromagnetism of Le-Bellac and L{\'e}vy-Leblond. Due to the presence of a dimensionless, gauge-invariant scalar field in the Galilean multiplet of the gauge-field, we find that at the quantum level an infinite number of couplings is generated. We explain how to handle the quantum corrections systematically using the background field method. Due to a non-renormalization theorem, the beta function of the gauge coupling is found to vanish to all orders in perturbation theory, leading to a continuous family of fixed points where the non-relativistic conformal symmetry is preserved.}

\begin{document}

\maketitle

\section{Introduction}
\label{sec:intro}
\input{sections/Introduction}

\section{Classical Aspects and Symmetries}
\label{sec:preliminaries}
\input{sections/Preliminaries}

\section{Non-renormalization Theorems}
\label{sec:nonrenorm}
\input{sections/nonrenorm}

\section{RG in Scalar Galilean Electrodynamics}
\label{sec:quantumGED}
\input{sections/RG_SGED}

\section{Discussion and Outlook}
\label{sec:discussion}
\input{sections/discussion}

\acknowledgments
We would like to thank Niayesh Afshordi, Igal Arav,  Eric Bergshoeff, Eyal Cornfeld, Jaume Gomis, Jelle Hartong, Matthijs Hogervorst, Zohar Komargodski, Niels Obers, Yaron Oz, Avia Raviv-Moshe and Shimon Yankielowicz for useful discussions.
SC would like to thank the organizers and participants of the workshops ``Beyond Lorentzian Geometry'' in Edinburgh and ``Gauge Theories and Black Holes'' at the Weizmann Institute, where this work was presented, for the many valuable comments. KTG is grateful for the hospitality of the Perimeter Institute and the Institute of Physics at the University of Amsterdam where part of this research was conducted. ZY is grateful for the hospitality of the Niels Bohr Institute. 
SC is supported by ERC consolidator grant QUANTIVIOL awarded to Ben Freivogel. 
LD is partially supported by INFN Iniziativa Specifica ST\&FI. This research is supported in part by Perimeter Institute for Theoretical Physics.
KTG acknowledges financial support from the Deutsche Forschungsgemeinschaft (DFG, German Research Foundation) under Germany's Excellence Strategy through the W\"urzburg-Dresden Cluster of Excellence on Complexity and Topology in Quantum Matter -- ct.qmat (EXC 2147, project--id 390858490) as well as the Hallwachs-R\"ontgen Postdoc Program of ct.qmat. 
Research at Perimeter Institute is supported in part by the Government of Canada through the Department of Innovation, Science and Economic Development Canada and by the Province of Ontario through the Ministry of Colleges and Universities.

\appendix

\section{Schr\"{o}dinger Symmetry Generators}\label{app:Gens}
\input{sections/appGens}

\section{Integrals and Regularization}\label{app:Integrals}
\input{sections/appIntegrals}

\section{Feynman Diagrams}\label{app:diagramsGED}
\input{sections/diagramsGED}

\section{Arbitrary Scalar Normalization}\label{app:sGEDCM}
\input{sections/sGEDCM}

\bibliographystyle{JHEP}
\bibliography{sGED}
\end{document}

%% file: sections/Introduction.tex
It has long been appreciated that symmetry principles play a major role in our understanding of physical systems. Symmetries involving transformations of space and time are particularly powerful and lead to various constraints on the dynamics of quantum field theories (QFTs). An important example is that of the conformal symmetry: in many interesting cases, this symmetry provides a handle on the strongly coupled regime of QFTs, leading to applications ranging from critical phenomena to string theory, see \eg \cite{Poland:2018epd}.

In the context of condensed matter theory, it is often the case that systems are governed at long distances by emergent symmetries that are different from the symmetries of the underlying microscopic theory. There is a-priori no reason why such effective symmetries should be relativistic, \ie why they should include the Lorentz symmetry group. As a consequence, many models have been proposed for condensed matter systems which are non-relativistic. In many cases we expect the system to be invariant under rotations, but additional symmetries could further restrict the dynamics. In particular, at criticality the system is expected to gain a scaling symmetry, which can treat space and time differently in the absence of a boost symmetry. For example, systems invariant under the Lifshitz group including an anisotropic scaling of space and time $t \rightarrow \lambda^z t$, $x^i \rightarrow \lambda x^i$, where $z$ is the dynamical critical exponent, were suggested as a potential explanation of the linear resistivity of strange metallic phases, see, \eg \cite{mcgreevy2010pursuit,Hartnoll:2009ns}.

A particularly interesting case is that of systems which besides rotations and scaling (with $z=2$) also enjoy an invariance under Galilean boosts and under a certain special conformal transformation. This symmetry is called Schr\"odinger symmetry, and it can be thought of as a non-relativistic analogue of the conformal symmetry \cite{Hagen:1972pd,Mehen:1999nd,Nishida:2007pj}. Though less powerful than its relativistic counterpart, this is arguably the most promising candidate to provide a starting point for a non-relativistic bootstrap program, and some first steps in that direction have been made in  \cite{Golkar:2014mwa,Goldberger:2014hca,Pal:2018idc,Gubler:2015iva,Nishida:2007pj}. A notable example governed by the Schr\"odinger symmetry is that of scattering of non-relativistic 3+1 dimensional spin-1/2 fermions in the infinite $S$-wave scattering length limit, also known as fermions at unitarity, see, \eg \cite{Mehen:1999nd,Nishida:2007pj,SonWingate}. This model has applications ranging from scattering in nuclear physics (characterized by an accidentally large scattering length) to experimentally-tunable ultra-cold atomic systems (see, \eg the introduction of \cite{Goldberger:2014hca} for a more detailed list of the relevant references). Field-theoretic descriptions can be given in terms of Schr\"odinger scalar fields driven to criticality by cubic or quartic interactions \cite{Nishida:2006br,Nishida:2006eu,Nikolic:2007zz}.

At low energies, not only can symmetries emerge, but even the effective field content of theories can differ from that of the underlying microscopic constituents. In condensed matter physics, one often encounters effective theories with emergent gauge fields. It is therefore natural to ask whether it is possible to incorporate interactions with gauge fields in non-relativistic effective field theories, and possibly find new fixed points with Schr\"odinger symmetry. In 2+1 dimensions, an action for gauge fields with the right symmetry is the Chern-Simons (CS) term. Indeed, another notable example of Schr\"odinger-invariant quantum field theory, which describes anyons, is obtained by coupling a CS gauge field to a Schr\"odinger scalar \cite{Hagen:1983rp, Hagen:1984mj, Jackiw:1990mb, Bergman,Nishida:2007pj,Doroud:2015fsz,Doroud:2016mfv}.

A different type of Schr\"odinger-invariant gauge theory, that can be defined in any number of space dimensions, has been proposed in the literature \cite{le1973galilean,Santos:2004pq,Festuccia}. This theory can be obtained from two different non-relativistic limits of Maxwell's equations, known as the electric and magnetic limits \cite{le1973galilean}, combined in a Galilean-invariant Lagrangian using auxiliary fields \cite{Santos:2004pq}.\footnote{In fact, the theory obtained in this way has a larger, infinite-dimensional spacetime symmetry group  \cite{Festuccia}, including independent time and space dilatations (for symmetries of the electric and magnetic limits see, \eg \cite{Bagchi:2014ysa,Duval:2009vt}). However, in this paper we are interested in the theory of the gauge field coupled to a Schr\"odinger matter field and that reduces the symmetry to the Schr\"odinger group.} It can also be derived from a non-relativistic limit of the theory of a gauge field and a real scalar field, and from a null reduction of relativistic Maxwell theory in one higher dimension \cite{Santos:2004pq, Bergshoeff:2015sic, Festuccia}. 
We refer to this theory as \emph{Galilean electrodynamics} (GED).
Previous studies of GED considered classical aspects of the theory, in particular its spacetime symmetries.
In this paper we initiate a study of the quantum mechanical properties of GED coupled to a Schr\"odinger scalar. The goal of finding new non-relativistic critical points motivates us to consider the case of 2+1 dimensions, in which the gauge coupling is marginal according to the $z=2$ scaling. We compute the beta functions of all couplings involved, and study their zeroes, indeed finding novel examples of Schr\"odinger-invariant quantum field theories.

The GED Lagrangian \cite{Santos:2004pq,Festuccia} is constructed using the gauge field $(a_t,a_i)$ together with an additional scalar field $\vp$ in the same Galilean multiplet, which is both dimensionless (in an anisotropic $z=2$ scaling sense) and inert under all the symmetries of the problem.\footnote{As we will show later, the additional scalar $\vp$ contributes to the Galilean boost transformations of the other gauge field components.} Without coupling to matter this theory does not contain propagating degrees of freedom. This is due to the fact that the velocity of light is infinite in non-relativistic theories. Therefore, similarly to CS theories with non-relativistic matter \cite{Bergman}, the gauge fields serve as instantaneous mediators of interactions between the matter fields. To introduce propagating degrees of freedom into the theory, we couple it to a scalar matter field $\sigma$ with a Schr\"odinger-like action. Together, the action reads\footnote{Here and in the following, we use the subscript sGED to denote GED coupled to a scalar field.}
\begin{equation} \label{intro:SsGED}
\begin{split}
\hspace{-5pt} S^{(0)}_\text{sGED} = \int dt& \, d^2 \mathbf{x} \biggl\{  \frac{1}{2} \dot\vp^2 + E^i \p_i \varphi - \frac{1}{4} f^{ij} f_{ij}
 \\ &\hspace{50pt} +\frac{i}{2} \bigl( \overline{\sigma} D_t \sigma - \sigma D_t \overline{\sigma} \bigr) - \frac{1}{2M} D_i \overline{\sigma} D^i \sigma
\biggr\}\,,
\end{split}
\end{equation}
where the spatial field-strength and electric field are defined as $f_{ij}\equiv\partial_i a_j -\partial_j a_i$ and $E_i\equiv\del_t a_i-\del_i a_t$, respectively, and the covariant derivatives are $D_{t} = \partial_{t} - i\, e\, q\, a_{t}$ and $D_{i} = \partial_{i} - i\, e\, q \, a_{i}$ with $e$  the gauge coupling and $q=+1$ $(-1)$ the charge of $\sigma$ ($\bar{\sigma}$).
The combination $M\equiv\Omega-e\vp$ encodes the minimal coupling of the field $\vp$ to the matter field through a shift of its mass parameter  $\Omega$. We will sometimes refer to the parameter $M$ as the covariantized mass/gap for reasons that will become clear in section \ref{sec:GED} when we discuss the relation of our model to null reduction.
Here the coefficient of the scalar spatial kinetic term $(2M)^{-1}$ should be understood as a series expansion in $e \varphi$, encapsulating infinitely many couplings.

The existence of the dimensionless scalar $\vp$ allows to construct infinitely-many classically-marginal operators, whose couplings turn out to have nonzero beta functions. We add to the GED action \eqref{intro:SsGED} the following action
\begin{equation} \label{intro:SsGED1}
\begin{split}
\hspace{-5pt} \Delta S_\text{sGED} = \int dt& \, d^2 \mathbf{x} \biggl\{ \mathcal{J}[M]  \, \p_i M \p^i M \, \overline{\sigma} \sigma - \frac{\lambda}{4}  \mathcal{V}[M]   \, (\overline{\sigma} \sigma)^2
- \mathcal{E} [M] \bigl( \p_i \p^i M - e^2  \, \overline{\sigma} \sigma \bigr) \, \overline{\sigma} \sigma
\biggr\}\,,
\end{split}
\end{equation}
including a subset of those couplings.
Similarly to the matter kinetic term discussed above, the couplings $\mathcal{J}[M]$, $\mathcal{V}[M]$ and $\mathcal{E}[M]$ should be understood as a series expansion in $e \varphi$, \eg $\mathcal{J}[M]=\mathcal{J}_0+e\mathcal{J}_1\vp+e^2\mathcal{J}_2 \vp^2+\ldots$. This subset of marginal couplings is closed under quantum corrections, namely, they mix among themselves but do not generate additional marginal couplings. In the process of evaluating the beta functions, we discuss the regularization of loop integrals in this non-relativistic setup.

We prove a set of non-renormalization theorems which allow us to deduce that the electromagnetic coupling $e$ does not run at any loop order.
We explain how to consistently evaluate the beta function(al)s for the infinitely many couplings $\mathcal{J}[M]$, $\mathcal{V}[M]$ and $\mathcal{E}[M]$ using a background-field method, similar to the calculation of the beta function in two-dimensional $\sigma$-models, see \eg \cite{ketov2013quantum}, and obtain at the one-loop order
\begin{align}
\begin{split}\label{intro:betasGED}
\beta_{\mathcal{J}[M]}& =\frac{e^2}{2\pi}\left(\mathcal{J}'[M]+\frac{5}{8 M^4}\right),\\
\beta_{\lambda \mathcal{V}[M]}& = \frac{e^4}{2\pi}\left( 4 \mathcal{J}[M] +\frac{5}{8 M^3}\right)
+\frac{\lambda \, e^2}{2\pi}\left(2 \mathcal{V}'[M] + \frac{\mathcal{V}[M]}{2 M}\right) + \frac{\lambda^2}{4\pi} M \mathcal{V}[M]^2\,,
\\
\beta_{\mathcal{E}[M]} & =\frac{e^2}{2\pi}\left(\mathcal{E}'[M] +\frac{1}{4 M^3}\right).
\end{split}
\end{align}
Fixed-points of the theory can be obtained by setting to zero the beta function(al)s above. These fixed point are only reliable when $e$ is very small, since they were obtained at one-loop, but their existence is robust and their location can be systematically corrected order by order in perturbation theory (\ie they are not due to beta functions identically vanishing at leading order).
We find that the theory has a manifold of fixed points labeled by $e$ and the constants of integration which appear when solving the differential equations for the vanishing of the beta functionals \eqref{intro:betasGED}. We demonstrate that the theories obtained at these fixed points are Schr\"odinger invariant and discuss their properties.

\sloppy

The appearance of conformal manifolds, \ie continuous families of fixed points parametrized by exactly marginal couplings, is rather surprising. In the context of ordinary, relativistic CFTs, conformal manifolds are ubiquitous in the presence of supersymmetry \cite{Leigh:1995ep, Green:2010da}, but it is much harder to find examples in non-supersymmetric theories in more than two spacetime dimensions.\footnote{See \cite{Bashmakov:2017rko, Behan:2017mwi, Hollands:2017chb, Sen:2017gfr} for a discussion of the constraints that need to be satisfied in order for a conformal manifold to exist.} Examples can be found by giving up unitarity, \eg in fishnet theories \cite{Gurdogan:2015csr,Grabner:2017pgm}, or by giving up locality, \eg in the context of boundary CFTs \cite{Herzog:2017xha,DiPietro:2019hqe,Herzog:2019bom}. Another example was found recently in a certain large $N$ vector model \cite{Chai:2020zgq}. Here we see that giving up Lorentz symmetry can also lead to conformal manifolds of the non-relativistic type, \ie obeying Schr\"odinger symmetry. Another non-relativistic example, albeit in a supersymmetric setup, appeared in \cite{Arav:2019tqm}, which found a line of fixed points with exact Lifshitz scale invariance. Ultimately, the existence of the continuous family of fixed points in our theory is due to the non-renormalization theorems, and it can be traced back to the fact that only processes that conserve the particle number are allowed in the non-relativistic limit, severely restricting the possible quantum corrections.

\fussy

The paper is organized as follows. In section \ref{sec:preliminaries}, we present basic facts about the Schr\"odinger symmetry group as well as the action, equations of motion, symmetries and currents for GED coupled to a scalar field. In section \ref{sec:nonrenorm}, we explain how our model gives rise to non-renormalization theorems which prevent the running of certain couplings.
In section \ref{sec:quantumGED}, we present the Feynman rules and perform the perturbative analysis. We conclude with a discussion of our results in section \ref{sec:discussion}. In appendix \ref{app:Gens}, we review how generators of the Schr\"odinger group can be constructed from the stress tensor and particle number current. In appendix \ref{app:Integrals}, we detail the technique we used to evaluate the integrals in our calculation of Feynman diagrams.
In appendix \ref{app:diagramsGED}, we list the results for the Feynman diagrams relevant for the calculations in section \ref{sec:quantumGED}.
Finally, in appendix \ref{app:sGEDCM} we clarify the role played by the field redefinition presented in section \ref{sec:preliminaries} in the renormalization process. 

%% file: sections/Preliminaries.tex
In this section, we review a number of preliminary ingredients required for understanding the analysis of this paper.
We start by reviewing some basic facts about the Schr\"{o}dinger symmetry group.
We then turn to the theory of \emph{Galilean electrodynamics} (GED) coupled to matter which we have chosen in this paper to be a single charged Schr\"{o}dinger scalar field with bosonic statistics.\footnote{There is no spin-statistics theorem for non-relativistic theories. Hence, the Schr\"{o}dinger scalar could in principle satisfy either bosonic or fermionic statistics. However, some of the interaction terms we include below, \eg $(\bar \sigma \sigma)^2$, would vanish for fermionic statistics. We comment further on this point in the discussion in section \ref{sec:discussion}.\label{statfoot}}
We refer to the theory with the scalar as \emph{scalar Galilean electrodynamics} (sGED).
We systematically classify all the marginal interaction terms for this theory and derive the equations of motion.
The theory of sGED enjoys a  Schr\"{o}dinger symmetry and we compute the associated conserved currents.
The quantum properties of sGED are explored later in sections \ref{sec:nonrenorm} and \ref{sec:quantumGED}.

\subsection{Schr\"{o}dinger Symmetry}
We work in $d+1$ spacetime dimensions, denoting time by $t$ and spatial coordinates by $\mathbf{x}$\,, with components $x^i$, $i = 1, \ldots, d$. The Schr\"{o}dinger symmetry is a non-relativistic analogue of the conformal symmetry obeyed, \eg by the free Schr\"{o}dinger equation. Its algebra consists of time translations (infinitesimally $\delta t=\xi^t$), space translations ($\delta x^i = \xi^i$), space rotations ($\delta x^i = \omega^i{}_j\, x^j$, with $\omega_{ij}$ anti-symmetric), Galilean boosts ($\delta t=0$, $\delta x^i =v^i t$), anisotropic $z=2$ dilatations ($\delta t = 2\lambda t$, $\delta x^i= \lambda x^i$) and special conformal transformations ($\delta t = - c t^2$, $\delta x^i = -ctx^i$), see, \eg \cite{Mehen:1999nd}. For later purposes it will be useful to also have the transformation of the derivatives under Galilean boosts
\begin{equation}\label{foot:derivboost}
\del_t \rightarrow \del_t - v^i \del_i\,, \qquad \del_i \rightarrow \del_i\,.
\end{equation}
We assign to these various transformations the following set of generators
\begin{align*}
	\hspace{2.5cm} \xi^t: & \quad \text{time translation} &&\hspace{-1.5cm} H\,, \\
	\xi^i: & \quad \text{space translations} &&\hspace{-1.5cm} P_i\,, \\
	\omega^{ij}: & \quad \text{spatial rotations} &&\hspace{-1.5cm} J_{ij}\,, \\ 
	v^i: & \quad \text{Galilean boosts} &&\hspace{-1.5cm} G_i\,, \\
	\lambda: & \quad \text{dilatations} &&\hspace{-1.5cm} D\,, \\
	c: & \quad \text{special conformal transformations} &&\hspace{-1.5cm} C\,.
\end{align*}
The generators also admit a central extension in terms of a mass associated with an extra generator $\MN$. This generator is related to particle number conservation (up to a constant pre-factor). The generators satisfy the following algebra (see, \eg \cite{Jensen:2014aia})\footnote{To obtain a Hermitian basis of generators, similar to those used in, \eg \cite{Nishida:2007pj,Son:2008ye}, we could redefine all generators by a factor of $-i$.}
\begin{equation}\label{eq:Schrodinger}
\begin{split}
&[J_{ij},J_{kl}]= \delta_{ik} J_{jl}+\delta_{jl}J_{ik}-\delta_{il}J_{jk}-\delta_{jk}J_{il},
\\
&[J_{ij},P_{k}]= \delta_{ik} P_{j}-\delta_{jk}P_{i}, \qquad
[J_{ij},G_{k}]= \delta_{ik} G_{j}-\delta_{jk}G_{i},
\\
& [D,H] = -2  H,\quad  [D,P_i]=- P_i, \quad [D,G^i]= G_i, \quad [D,C]=2 C,
\\
&
[P_i,G_j]=-\delta_{ij} \MN, \quad [H,C]=D,\quad  [H, G_i]  = -P_i, \quad [P_i, C]  = -G_i.
\end{split}
\end{equation}

The various symmetry generators above can be expressed in terms of the energy density $T^t{}_{t}$, the energy flux $T^i{}_{t}$, the momentum density $T^t{}_{i}$, the momentum flux $T^i{}_{j}$, the $U(1)$ mass density $J_{m}^{t}$ and the mass flux $J_m^i$ (also often referred to as the mass current). We summarize these expressions for the symmetry generators in appendix \ref{app:Gens}; also see \cite{Nakayama:2009ww,Nakayama:2013is}. The stress tensor and mass current satisfy a number of conditions following from the Schr\"odinger symmetry (see also \cite{Jackiw:1990mb,Arav:2016xjc} and appendix A of \cite{Nakayama:2009ww}). First, invariance under space and time translations implies that the energy momentum tensor is conserved,
\begin{equation}\label{eq:consT}
\del_t T^{t}{}_{t}+\del_i T^{i}{}_{t}=0, \qquad \del_t T^t{}_i+\del_j T^j{}_i=0\,.
\end{equation}
Invariance under spatial rotations implies that the spatial components of the stress tensor are symmetric, \ie $T^{ij}=T^{ji}$\,.
Invariance under Galilean boosts implies that the momentum density and mass flux are equal to each other
\begin{equation} \label{eq:TJWard}
T^{ti}=J_m^i\,.
\end{equation}
Invariance under the global $U(1)$ symmetry associated with the central extension $\MN$ implies that the associated current is conserved
\begin{equation}\label{eq:Jconservation}
\del_t J_{m}^{t} + \del_i J_m^i=0\,.
\end{equation}
Finally, conformal invariance implies that the stress tensor can be improved such that\footnote{Invariance under dilatations only implies that $2T^t{}_t+T^i{}_i=\del_t \mathcal{S} + \del_j W^j$. See appendix A of \cite{Nakayama:2009ww} or section 9 of \cite{Nakayama:2013is} for further details.\label{THEEEFOOTNOTE}}
\begin{equation}\label{eq:Ttrace}
2T^t{}_t+T^i{}_i=0 \,.
\end{equation}
The above identities can be derived from the Noether theorem, or alternatively by placing the theory on a curved background with a Newton-Cartan geometry and taking variations with respect to the background fields, see, \eg  \cite{Jensen:2014wha,Arav:2016xjc}.

\subsection{Scalar Galilean Electrodynamics} \label{sec:GED}
From now on, we will focus our attention on $(2+1)$-dimensions. The GED theory is defined in terms of a $U(1)$ gauge field,  $(a_t\,, a_i)$, $i\in\{1,2\}$, together with an additional scalar field $\varphi$ in the gauge multiplet. The field $\varphi$ is invariant under both $U(1)$ gauge transformations and Galilean boosts. Let us start by clarifying the role of this additional scalar $\vp$. The $U(1)$ gauge transformations are given by
\begin{equation}\label{eq:GaugeCStrns}
    a_t \rightarrow a_t + \p_t \varepsilon\,,
    \qquad
    a_i \rightarrow a_i + \p_i \varepsilon\,,
    \qquad
    \vp \rightarrow \vp\,.
\end{equation}
At first sight, we expect the gauge fields to transform under boosts in a similar way to the temporal and spatial derivatives, see eq.~\eqref{foot:derivboost}, \ie
\begin{align} \label{eq:Gboostphi0}
\begin{split}
    a'_t (t', \mathbf{x}')   & = a_t (t, \mathbf{x}) - v^i a_i (t, \mathbf{x})\,, \\
    a'_i (t', \mathbf{x}') & = a_i (t, \mathbf{x})\,,
\end{split}
\end{align}
in such a way that covariant derivatives of charged matter fields will transform covariantly under Galilean boosts.

Under the transformation rules \eqref{eq:GaugeCStrns} and \eqref{eq:Gboostphi0}, the only quadratic gauge and Galilean boost-invariant action is a spatial Maxwell-like kinetic term of the form
\begin{equation}\label{eq:SMaxwell}
	S_{\text{GM}} = \int dt \, d^2 {\bf x} \, \lr \, - \frac{1}{4} f_{ij} f^{ij} \rr \,,
\end{equation}
where here and in the following we define the gauge invariant field strengths as
\be \label{eq:Ef}
    E_i \equiv \p_t a_i - \p_i a_t\,,
        \qquad%
    f_{ij} \equiv \p_i a_j - \p_j a_i\,,
\ee
and refer to $E_i$ as the electric field and to $f_{ij}$ as the magnetic field.
The action \eqref{eq:SMaxwell} depends only on the magnetic field.
However, once the additional scalar $\varphi$ is included, a Galilean and gauge-invariant action can be constructed, which will involve both the electric and magnetic fields.\footnote{Note that in 2+1 dimensions there is also a parity-odd Galilean-invariant term that can be added to the action \eqref{eq:SMaxwell}, namely the Chern-Simons term $\epsilon^{\mu\nu\rho}a_\mu \del_\nu a_\rho$ with $\mu,\nu,\rho \in\{t,1,2\}$. The spatial Maxwell+CS gauge theory can then be coupled to a Schr\"odinger scalar with the quartic interaction $\lambda (\bar \sigma \sigma)^2$. Up to equations of motion, however, the addition of the spatial Maxwell term is equivalent to a shift in the quartic coupling. Therefore, this addition can be absorbed by a field redefinition, and the beta function for $\lambda$ follows straightforwardly from the calculation without the extra spatial Maxwell term.  The theory of a CS gauge field coupled to a Schr\"odinger scalar with quartic interaction is discussed in the context of anyons \cite{Bergman,Nishida:2007pj}.}

The theory obtained in this way is referred to as \emph{Galilean Electrodynamics} (GED) and was previously studied in \cite{Festuccia,Santos:2004pq} as a Lagrangian constructed in order to combine the electric and magnetic limits of Le Bellac and L{\'e}vy-Leblond \cite{le1973galilean}. The theory is described by the following kinetic action for the gauge fields
\begin{equation} \label{eq:GED}
	S_\text{GED} = \int dt \, d^2 {\bf x} \lr \frac{1}{2} \, \dot \vp^2  + E^i \p_i \varphi - \frac{1}{4} f^{ij} f_{ij} \rr\,,
\end{equation}
where the dot stands for a derivative with respect to the time coordinate.\footnote{It turns out that the GED action is the unique quadratic action (at second order in derivatives) which can be constructed with the addition of a gauge-invariant scalar field $\vp$ to the gauge multiplet, and which is invariant under gauge, Galilean boosts and a $z=2$ anisotropic scaling symmetry.}
To make the action \eqref{eq:GED} invariant under Galilean boosts, the Galilean boost transformations need to be modified by $\vp$-dependent terms and are given by \cite{Festuccia}
\begin{align} \label{eq:Gboostphi}
\begin{split}
    a'_t (t', \mathbf{x}')   & = a_t (t, \mathbf{x}) - v^i a_i (t, \mathbf{x}) - \tfrac{1}{2} v^i v_i \, \varphi (t, \mathbf{x})\,, \\[2pt]
    a'_i (t', \mathbf{x}') & = a_i (t, \mathbf{x}) + v_i \, \varphi (t, \mathbf{x})\,,\quad \\[2pt]
    \varphi' (t', \mathbf{x}') & = \varphi (t, \mathbf{x})\,.
\end{split}
\end{align}
Note that the spatial Maxwell term \eqref{eq:SMaxwell} by itself is no longer invariant under the Galilean boosts in eq.~\eqref{eq:Gboostphi}.\footnote{Similarly, the Chern-Simons term $\epsilon^{\mu\nu\rho}a_\mu \del_\nu a_\rho$ is no longer invariant under Galilean boosts after the addition of the $\varphi$ dependent contributions to the transformation laws \eqref{eq:Gboostphi}.}

The theory in eq.~\eqref{eq:GED} was originally derived as a null-reduction of a relativistic Maxwell theory in 3+1 dimensions \cite{Santos:2004pq,Festuccia}. Explicitly, one reduces the Maxwell action along the $x^+$ null direction of the following coordinate system: $ds^2 = 2 dx^+ dx^-+ (dx^i)^2$, where the four-dimensional gauge field $\mathcal{A}$ is related to the GED fields as follows: $\mathcal{A}_I\equiv(\mathcal{A}_+,\mathcal{A}_-,\mathcal{A}_i)=(\vp,a_t,a_i)$. The gauge fields are taken to be independent of the $x^+$ coordinate, while the $x^-$ coordinate plays the role of the time $t$ in the (2+1)-dimensional theory. The Schr\"odinger scalar, which we couple to GED below, can also be obtained by null-reducing a relativistic complex scalar field according to $\Phi(x^+,x^-,x^i) = e^{i \Omega x^+}\sigma(x^-,x^i)$. In fact, this reduction can be seen as  the origin of the recurring combination $M\equiv \Omega-e\vp$ (see below), which is simply the covariant derivative in the $x^+$ direction.\footnote{Let us further note that applying the null reduction to the $\theta$-term in Maxwell's theory in four dimensions, \ie $\epsilon^{\mu\nu\rho\sigma} F_{\mu\nu} F_{\rho\sigma}$ results in a $(2+1)$-dimensional action proportional to
$\epsilon^{ij} \bigl( 2 \, \p_i \varphi \, E_j - \p_t \varphi \, f_{ij} \bigr)$. Unlike the relativistic $\theta$-term, this term is a total derivative of a gauge-invariant term and hence it can be ignored for spacetime manifolds without boundaries.}

The theory \eqref{eq:GED} is invariant under anisotropic $z=2$ scaling. In our analysis below, it will be convenient to keep track of the anisotropic scaling dimensions of various fields and coordinates:
\begin{align}
    [t] &= - 2\,, &
    [x^i] &= - 1\,, &
	[a_t] &= 2\,, &
	[a_i] &= 1\,, &
	[\varphi] &= 0\,,
\end{align}
such that Lagrangian densities have a scaling dimension of $[\mathcal{L}]=d+2$ and actions are dimensionless.

One may consider self-interactions in the gauge sector, around the fixed point defined by the free theory in eq.~\eqref{eq:GED}. Up to integration by parts, the independent marginal interactions are
\begin{equation} \label{eq:GEDint0}
\begin{split}
	\mathcal{G}_1 [\varphi] \, \varphi \lr \frac{1}{2} \, \dot\vp^2 + E^i \p_i \varphi - \frac{1}{4} f^{ij} f_{ij} \rr
+\mathcal{G}_2 [\varphi] \, \bigl( \p^i \varphi \, \p_i \varphi \bigr)^2
\\
+\mathcal{G}_3 [\varphi] \, \varphi \, \p^i \p^j \varphi \, \p_i \p_j \varphi +
\mathcal{G}_4 [\varphi] \,  \p^i \p^j \varphi \, \p_i \varphi \, \p_j \varphi\,,
\end{split}
\end{equation}
where, $\mathcal{G}_i [\varphi]$\,, $i=1,\ldots,4$ are functions of $\varphi$, which should be understood as a Taylor expansion in $e\varphi$\,.
In section \ref{sec:nonrenorm}, we prove a non-renormalization theorem which states that none of the interaction terms in \eqref{eq:GEDint0} will be generated along the renormalization group (RG) flow at any loop order when we couple the free GED theory \eqref{eq:GED} to a Schr\"odinger scalar.
Therefore, we will focus on the minimal setup with the GED action defined in eq.~\eqref{eq:GED}.

The theory \eqref{eq:GED} does not possess propagating degrees of freedom. This can be seen by exploring the structure of the poles in the gauge field propagator, or alternatively by a Dirac constraint analysis, see, \eg \cite{Banerjee:2019axy}.\footnote{We explicitly demonstrate this later in section \ref{sec:quantumGED} by showing that the pole in the gauge field propagator does not have any frequency dependence. As a consequence, in position space, the propagator is instantaneous in time.} To introduce some propagating degrees of freedom, we couple the GED theory to a Schr\"{o}dinger scalar field $\sigma$. The Schr\"odinger field transforms under $U(1)$ gauge as
\begin{equation}\label{eq:GboostU1sigma}
\sigma \rightarrow \sigma \, e^{i e \varepsilon}\,,
\end{equation}
and under Galilean boosts as
\begin{equation}
\label{eq:Gboostsigma}
    \sigma' (t', \mathbf{x}')  = \exp \! \ls i \Omega \lr \frac{1}{2} v^2 t + v_i x^i \rr \rs \sigma (t, \mathbf{x})\,,
\end{equation}
where $\Omega$ is a constant mass parameter. In the absence of the scalar field $\vp$, the above transformation laws leave the Schr\"odinger action
\begin{align}
 \label{eq:Ssigma0}
    S_\sigma^{(0)}  =  & \int dt \, d^2 {\bf x} \ls \frac{i}{2} \bigl( \overline{\sigma} D_t \sigma - \sigma D_t \overline{\sigma} \bigr) - \frac{1}{2\Omega} \, D_i \overline{\sigma} \, D_i \sigma \rs
\end{align}
invariant, where the covariant derivatives are defined as usual by
\begin{equation}
\begin{split}\label{covder}
    D_t \sigma & = \bigl( \p_t - i e a_t \bigr) \, \sigma\,,
        \qquad
    D_i \sigma = \bigl( \p_i - i e a_i \bigr) \, \sigma\,, \\
    D_t \overline{\sigma} & = \bigl( \p_t + i e a_t \bigr) \, \overline{\sigma}\,,
        \qquad
    D_i \overline{\sigma} = \bigl( \p_i + i e a_i \bigr) \, \overline{\sigma}\,.
\end{split}
\end{equation}
With the addition of the scalar $\vp$\,, however, the action \eqref{eq:Ssigma0} is no longer invariant. This situation can be remedied by the following change \cite{Festuccia}:
\begin{align}
 \label{eq:Ssigma1}
    S_{\sigma,\text{GED}}^{(0)}  =  & \int dt \, d^2 {\bf x} \ls \frac{i}{2} \bigl( \overline{\sigma} D_t \sigma - \sigma D_t \overline{\sigma} \bigr) - \frac{1}{2M} \, D_i \overline{\sigma} \, D_i \sigma \rs,
\end{align}
where the parameter
\be \label{eq:MfromOmega}
	M \equiv \Omega - e \varphi
\ee
``covariantizes" the mass parameter $\Omega$ in the denominator of the scalar spatial kinetic term in eq.~\eqref{eq:Ssigma0}. This particular modification of the scalar action is required in order to maintain boost invariance with the transformation rules in eq.~\eqref{eq:Gboostphi}. The coefficient of the scalar spatial kinetic term involves a negative power of $M$ and should be understood as a series expansion in 
$e \vp$.\footnote{We could have started our analysis with the Lagrangian $ i\, M\bigl( \overline{\sigma} D_t \sigma - \sigma D_t \overline{\sigma} \bigr) -  D_i \overline{\sigma} D^i \sigma$ which looks simpler at the classical level because $\vp$ appears linearly. Even with this choice, however, infinitely many quantum corrections will be generated, which together are equivalent to negative powers of $M$. This is because the two Lagrangians are related by a field redefinition and they are equivalent at the quantum level. \label{foot:negativekinetic}}

The presence of the dimensionless scalar $\vp$, which is inert under Galilean boosts and gauge transformations, implies that the space of possible couplings allowed is actually much larger than those encapsulated in the $(2M)^{-1}$ term in equation \eqref{eq:Ssigma1}.
Keeping only marginal terms, we consider the most general action given by
\begin{align}\label{eq:totalaction}
\begin{split}
    S_\text{sGED} & = \int dt \, d^2 \mathbf{x} \biggl\{ \frac{1}{2} \, \dot \vp^2 + E^i \p_i \varphi - \frac{1}{4} f^{ij} f_{ij}  \\[2pt]
	& + \mathcal{C}[M] \ls \frac{i}{2} \bigl( \overline{\sigma} D_t \sigma - \sigma D_t \overline{\sigma} \bigr) - \frac{1}{2M} D_i \overline{\sigma} D_i \sigma \rs - \mathcal{P}[M, \p_i] \, \overline{\sigma} \sigma - \frac{1}{4} \,\lambda\, \mathcal{Q}[M] \, (\overline{\sigma} \sigma)^2 \biggr\}\,.
\end{split}
\end{align}
The functions $\mathcal{C}[M]$ and $\mathcal{Q}[M]$ are power series in $\varphi$\,, which encode the kinetic term of the matter scalar $\sigma$ as well as various interaction terms. Similarly, $\mathcal{P}(M, \p_i)$ is a power series in $M$\,, $\p_i M$ and $\p_i \p^i M$ containing exactly two spatial derivatives. We may further simplify the action \eqref{eq:totalaction} by performing a field redefinition,
\be\label{eq:redef}
	\sigma \rightarrow \frac{\sigma}{\sqrt{\mathcal{C} [M]}}\,,
\ee
which yields
\begin{align} \label{eq:sGEDaction}
\begin{split}
\hspace{-5pt} S_\text{sGED} = \int dt \, & d^2 \mathbf{x} \,  \biggl\{ \frac{1}{2} \dot\vp^2 + E^i \p_i \varphi - \frac{1}{4} f^{ij} f_{ij} + \frac{i}{2} \bigl( \overline{\sigma} D_t \sigma - \sigma D_t \overline{\sigma} \bigr) - \frac{1}{2M} D_i \overline{\sigma} D^i \sigma \\[2pt]
	& + {\mathcal{J}}[M] \, \p_i M \p^i M \, \overline{\sigma} \sigma - \frac{1}{4} \, \lambda \, {\mathcal{V}}[M] \, (\overline{\sigma} \sigma)^2 - \mathcal{E} [M] \bigl( \p_i \p^i M - e^2 \overline{\sigma} \sigma \bigr) \overline{\sigma} \sigma \biggr\} \,,
\end{split}
\end{align}
where the subscript sGED stands for \emph{scalar Galilean electrodynamics}.
Here, $\mathcal{J}[M]$, $\mathcal{V}[M]$ and $\mathcal{E}[M]$ are power series in $e \vp$ that encode the infinitely many couplings of the theory.
As we show later in section \ref{sec:quantumGED}, the inclusion of these couplings is forced upon us by the RG flow.
The action \eqref{eq:sGEDaction} is invariant under the full Schr\"odinger group as we demonstrate in the next subsection.\footnote{Ref.~\cite{Festuccia} carried out an extensive analysis of the symmetries of Galilean electrodynamics (without matter) in any dimension. However it was missed that in 2+1 dimensions the theory does enjoy a symmetry under the special conformal transformation of the Schr\"odinger group. We thank N. Obers and J. Hartong for discussions on this topic.} In appendix \ref{app:sGEDCM}, we show explicitly that the classical field redefinition \eqref{eq:redef} does not affect the properties of the theory at the quantum mechanical level.

The equations of motion for the action \eqref{eq:sGEDaction} are given by
\begin{subequations}\label{eq:sGEDeomat}
\begin{align}
 	\p_i \p^i \varphi + e \, \overline{\sigma} \sigma = &\,  0\,,\label{eq:eomvarphi}\\
\p_j f^j{}_{i} - \p_i \p_t \varphi   =  &\,  \tfrac{i\,e}{2M}  \bigl( \overline{\sigma} D_i \sigma - \sigma D_i \overline{\sigma} \bigr)\,,
\\
\begin{split}
-\p_t^2 \varphi - \p_i E^i   = &\,   \tfrac{e}{2M^2} D_i \overline{\sigma} D^i \sigma - \tfrac{e}{4} \lambda \mathcal{V}'[M] \bigl( \overline{\sigma} \sigma \bigr)^2 - e\, \p_i \p^i \bigl( \mathcal{E}[M] \, \overline{\sigma} \sigma \bigr)\\
	&   - e \left[ \p_i \mathcal{J}[M] \p^i M \, \overline{\sigma} \sigma + 2 \, \mathcal{J}[M] \p_i \bigl( \p^i M \, \overline{\sigma} \sigma \bigr) \right] \,,
\end{split}\\
\begin{split}
  i D_t \sigma + D_i \left(\tfrac{1}{2M}  D^i \sigma\right) =  &\,
- \mathcal{J}[M] \bigl( \p_i M \p^i M \bigr) \sigma
 + \bigl( \tfrac{1}{2} \lambda \mathcal{V}[M] - e^2 \mathcal{E}[M] \bigr) (\overline{\sigma} \sigma) \sigma \,,
\end{split}
\end{align}
\end{subequations}
where we have used eq.~\eqref{eq:eomvarphi} to simplify some of the other equations of motion.
Note that the operator with the coupling $\mathcal{E} [M]$ in the action \eqref{eq:sGEDaction} is proportional to the equation of motion
\eqref{eq:eomvarphi}. Such couplings do not contribute to the beta functions of other operators which do not vanish on-shell, as is indeed apparent from eq.~\eqref{intro:betasGED}; see, \eg the explanation around eqs.~(6.40)-(6.41) in \cite{Manohar:2018aog}.
As a consequence, our choice of basis for the marginal operators in eq.~\eqref{eq:sGEDaction} provides a significant simplification of the expressions for the running of the coupling constants computed in section \ref{sec:quantumGED}.

\subsection{Conserved Currents}
In this subsection, we demonstrate that the stress tensor and mass current associated with the theory \eqref{eq:sGEDaction} obey the identities \eqref{eq:consT}-\eqref{eq:Ttrace} and thus sGED is Schr\"odinger invariant at the classical level for any value of the couplings $\mathcal{J}[M]$, $\mathcal{V}[M]$ and $\mathcal{E}[M]$.
To find the energy-momentum tensor, we use Noether's theorem for a (global) infinitesimal spacetime translation which
acts on the fields of sGED as follows:
\begin{equation}
\begin{split}
	{\delta}_\xi a_t & = \xi^t \p_t a_t + \xi^j \p_j a_t \,,\qquad
    	{\delta}_\xi a_i  = \xi^t \p_t a_i + \xi^j \p_j a_i\,,
    \\
    	{\delta}_\xi \vp & = \xi^t \p_t \vp + \xi^j \p_j \vp\,,\qquad
	{\delta}_\xi \sigma  = \xi^t \p_t \sigma + \xi^i \p_i \sigma \,.
\end{split}
\end{equation}
These transformations are not gauge covariant, and therefore the associated energy-momentum tensor (before improvements) will not be manifestly gauge invariant. Instead, we can ensure gauge invariance of the energy-momentum tensor by defining it with respect to a transformation which is a mixture of a translation and a $U(1)$ gauge transformation, \ie
\begin{align}
    	\tilde{\delta}_\xi a_t & = \delta_\xi a_t + \p_t \varepsilon \,,
        		&%
    	\tilde{\delta}_\xi a_i & = \delta_\xi a_i + \p_i \varepsilon\,,
    		&%
     \tilde{\delta}_\xi \vp&=\delta_\xi \vp\,,
     &%
	\tilde{\delta}_\xi \sigma & = \delta_\xi \sigma  + i e \varepsilon \sigma\,.
\end{align}
Fixing $\varepsilon = - \xi^t a_t - \xi^i a_i$ yields the following manifestly gauge covariant transformation
\begin{align} \label{eq:translationxi}
    \hspace{-5pt}\tilde{\delta}_\xi a_t = - \xi^i E_i\,,
        \quad%
    \tilde{\delta}_\xi a_i = \xi^t E_i + \xi^j f_{ji}\,,
        \quad%
    \tilde{\delta}_{\xi} \varphi = \xi^t \p_t \varphi + \xi^i \p_i \varphi\,,
    \quad
    \tilde{\delta}_\xi \sigma = \xi^t D_t \sigma + \xi^i D_i \sigma\,.
\end{align}
Using the Noether procedure for the transformation $\tilde{\delta}_{\xi}$ in eq.~\eqref{eq:translationxi}, we find that the energy-momentum tensor is given by
\begin{subequations} \label{eq:emtGED}
\begin{align}
\tilde T^t{}_t  = &\,  \dot{\varphi}^2 + E_i \p^i \varphi + \frac{i}{2} \bigl( \overline{\sigma} D_t \sigma - \sigma D_t \overline{\sigma} \bigr) - \del_k (\mathcal{E}[M]\del^k M \bar \sigma \sigma)- \mathcal{L}\,,
\\		
\begin{split}
\tilde T^i{}_t   = &\, \dot{\varphi} \, E^i   - E_j f^{ij} - \tfrac{1}{2M} \bigl( D_t \overline{\sigma} D^i \sigma + D_t \sigma D^i \overline{\sigma} \bigr)
\\
  &\, + \bigl[ \p^i M \, \p_t + \p_t M \, \p^i \bigr] ( \mathcal{E}[M] \, \overline{\sigma} \sigma ) + 2 \mathcal{J}[M] \, \p^i M \, \p_t M \, \bigl( \overline{\sigma} \sigma \bigr)\,,
\end{split}
\\
\tilde T^t{}_i  =&\,  \dot{\varphi}\, \p_i \varphi  + f_{ij} \p^j \varphi + \frac{i}{2} \bigl( \overline{\sigma} D_i \sigma - \sigma D_i \overline{\sigma} \bigr)\,,
\\
\begin{split}
\tilde T^i{}_j  =&\,  \bigl( E_j \p^i \varphi + E^i \p_j \varphi \bigr) - f_{jk} f^{ik} - \tfrac{1}{2M} \bigl( D_j \overline{\sigma} D^i \sigma + D^i \overline{\sigma} D_j \sigma \bigr)
\\
&\, + 2 \mathcal{J}[M] \, \p^i M \, \p_j M \bigl( \overline{\sigma} \sigma \bigr)  + \bigl[ \p^i M \, \p_j + \p_j M \, \p^i \bigr] ( \mathcal{E}[M] \, \overline{\sigma} \sigma )
\\
&\, -  \delta^i{}_j \,\del_k (\mathcal{E}[M]\del^k M \bar \sigma \sigma) - \delta^i{}_j \mathcal{L}\,,
\end{split}
\end{align}
\end{subequations}
where $\mathcal{L}$ is the Lagrangian density associated with the action $S_\text{sGED}$ in \eqref{eq:sGEDaction}. Of course, these expressions
satisfy the conservation \eqref{eq:consT} by construction.
One can further verify that the energy-momentum tensor \eqref{eq:emtGED} satisfies
\be \label{eq:TttTii}
    2 \tilde{T}^t{}_t + \tilde{T}^i{}_i = 2 \p^i \mathcal{O}_i\,,
\ee
where
\be \label{eq:OiE}
	\mathcal{O}_i = \frac{1}{4M} \, \p_i \bigl( \overline{\sigma} \sigma \bigr) - \mathcal{E}[M] \, \p_i M \bigl( \overline{\sigma} \sigma \bigr) \,.
\ee
The energy-momentum tensor can then be improved to satisfy the trace condition \eqref{eq:Ttrace} by redefining
\begin{align} \label{eq:improveGEDT}
    	T^t{}_t & = \tilde T^t{}_t - \p^i \CO_i\,,
      		\qquad%
    	T^i{}_t = \tilde T^i{}_t + \p_t \CO_i\,,
		\qquad%
      	T^t{}_i = \tilde{T}^t{}_i\,,
    		\qquad
	T^i{}_j = \tilde{T}^i{}_j\,,
\end{align}
and this redefinition does not affect its conservation.

We proceed with the mass current associated with the infinitesimal global transformation
\begin{align} \label{eq:globalU1}
    \delta_\alpha \sigma &= i \,\Omega \,\alpha\, \sigma, \qquad%
    \delta_\alpha \overline{\sigma} = - i \,\Omega \,\alpha\, \overline{\sigma},
\end{align}
where we recall that $\Omega$ is the mass parameter which appears in the Galilean boost transformation of $\sigma$, and $\alpha$ is a constant real parameter.
The transformation \eqref{eq:globalU1} is a global transformation and is \emph{not} accompanied by a transformation of the gauge fields.
To derive the mass current associated with \eqref{eq:globalU1}, we apply again the Noether procedure.
As with the previous derivation of the energy-momentum tensor, we mix this global $U(1)$ transformation with a gauge $U(1)$ transformation, see eqs.~\eqref{eq:GaugeCStrns} and \eqref{eq:GboostU1sigma}, in this case with parameter $\varepsilon = - \varphi \alpha$. Doing this will ensure that the spatial current $J_{m}^{i}$ associated with this transformation equals $T^{ti}$, as implied by Galilean boost invariance, see eq.~\eqref{eq:TJWard}.
The complete infinitesimal transformation is then given by
\be \label{eq:tildedeltaepsilon}
	\tilde{\delta}_\alpha \sigma = i ( \Omega \, \alpha + e \, \varepsilon ) \sigma= i M \alpha \, \sigma\,,
		\qquad
	\tilde{\delta}_\alpha a_t = \p_t \varepsilon= - \alpha \, \p_t \varphi\,,
		\qquad
	\tilde{\delta}_\alpha a_i = \p_i \varepsilon= - \alpha \, \p_i \varphi\,.
\ee
The mass current associated with the transformation in eq.~\eqref{eq:tildedeltaepsilon} is
\be \label{eq:JmtGED}
	J^t_m  = -  \, (\p_i \varphi)^2 - M \, \overline{\sigma} \sigma\,, \qquad J_m^i = \dot{\varphi}\, \p_i \varphi  + f_{ij} \p^j \varphi + \frac{i}{2} \bigl( \overline{\sigma} D_i \sigma - \sigma D_i \overline{\sigma} \bigr)\,.
\ee
Note that $J_m^t$ and $J_m^i$ satisfy the conservation law \eqref{eq:Jconservation} as well as the identity \eqref{eq:TJWard} following from Galilean boost invariance. Using the equation of motion \eqref{eq:eomvarphi}, the mass density $J^t_m$ can be brought to the form of a total derivative $J_{m}^{t} = \frac{1}{e} \partial_i ( M \partial_i \vp )$. Note that this is the total derivative of a globally well-defined, gauge-invariant operator. As a consequence, any gauge-invariant local operator must have vanishing central charge $N=0$ and all local correlation functions are forced to the $N=0$ sector of the Schr\"odinger algebra. This can be understood as a consequence of the fact that the $U(1)$ number-conservation symmetry of the Schr\"odinger scalar is being gauged by the coupling to GED.

%% file: sections/nonrenorm.tex
In this section, we prove a set of non-renormalization theorems, applicable to abelian gauge theories coupled to a single Schr\"odinger scalar, in  arbitrary dimension.
The core of the argument relies on the fact that, in many cases, oriented loops of the single Schr\"odinger scalar vanish, see, \eg \cite{Bergman:1991hf,Klose:2006dd,Auzzi:2019kdd}.\footnote{We thank Igal Arav and Avia Raviv-Moshe for discussions on this point.} We will show that, as a consequence, in the sGED model, the electric coupling constant $e$ does not run at any loop order.

To understand this claim, let us examine the propagator of the Schr\"odinger scalar field. The quadratic part of the Schr\"odinger action is given by
\be \label{eq:Ssigma(0)2}
    S^{(0)}_\sigma = \int dt \, d^d {\bf x} \ls \frac{i}{2} \bigl( \overline{\sigma} \p_t \sigma - \sigma \p_t \overline{\sigma} \bigr) - \frac{1}{2\Omega} \, \p_i \overline{\sigma} \, \p_i \sigma \rs\,,
\ee
and so the propagator for the $\sigma$ field reads
\begin{align}\label{propiepsilon}
&\raisebox{-0.3\height}{\includegraphics[width=2.8in]{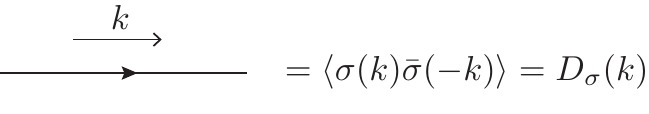}}
 = \frac{i}{\omega - \frac{{\bf k}^2}{2 \Omega}+i \epsilon}\,,
\end{align}
where $k \equiv (\omega\,, \mathbf{k})$. In what follows, we will use an arrow on the scalar line to indicate the flow of charge and an additional thinner arrow on top of the line to indicate the flow of momentum. In the above propagator, we are using the $+i \epsilon$ prescription which indicates that the propagator for the scalar will be time ordered, also allowing for a Wick rotation in the momentum integrals $\omega = i \omega_E$. The single frequency pole implies that the position space propagator is proportional to a Heaviside theta function
\be \label{eq:posspaceprop}
G (t-t', \mathbf{x}-\mathbf{x}')
 \equiv \int \frac{d\omega\, d^d {\bf k}}{(2\pi)^{d+1}}e^{-i\omega (t-t') +i \mathbf{k} \cdot (\mathbf{x}-\mathbf{x}')}\, D_\sigma(k)=
\theta (t - t') \, G_W (t-t', \mathbf{x}-\mathbf{x}'),
\ee
where $G_\text{W} (t-t', \mathbf{x}-\mathbf{x}')$ is the Wightman function defined by\footnote{The position space propagator is given by the time ordered expectation value
$G (t-t', \mathbf{x}-\mathbf{x}')=\langle0| T(\sigma (t, \mathbf{x}) \, \overline{\sigma} (t', \mathbf{x}')) |0\rangle=\theta(t-t')\langle0| \sigma (t, \mathbf{x}) \, \overline{\sigma}( t', \mathbf{x}')|0\rangle+\theta(t'-t)\langle0|  \, \overline{\sigma}( t', \mathbf{x}')\sigma (t, \mathbf{x})|0\rangle$, where in Galilean theories $\sigma(t,{\bf x})$ is expanded only in terms of annihilation operators and therefore the second part of this expression vanishes. For this reason we obtain that the time-ordered two-point function is proportional to $\theta(t-t')$. Stripping off the theta function then leads to the Wightman function $G_W(t-t', \mathbf{x}-\mathbf{x}')=\langle0| \sigma (t, \mathbf{x}) \, \overline{\sigma}( t', \mathbf{x}')|0\rangle$.}
\be \label{eq:posspaceprop2}
	G_\text{W} (t-t', \mathbf{x}-\mathbf{x}') = \langle 0 | \sigma (t, \mathbf{x}) \, \overline{\sigma} (t', \mathbf{x}') | 0 \rangle =  \biggl[ \frac{\Omega}{2\pi i (t - t')} \biggr]^\frac{d}{2} \exp \biggl( \frac{i}{2} \frac{\Omega | \mathbf{x} - \mathbf{x}' |^2}{t - t'} \biggr)\,.
\ee

\begin{figure}
\centering
~~~\begin{subfigure}[b]{0.4\linewidth}\includegraphics[width=\textwidth]{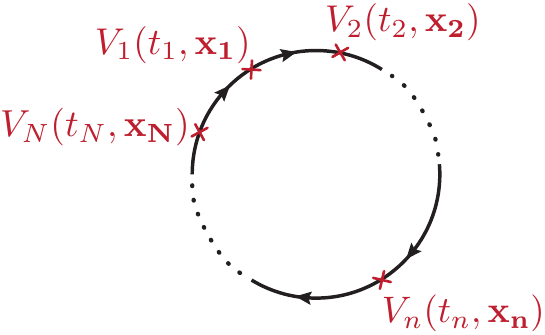}\vspace{4pt}
\caption{In position space.}\label{fig:nonrenormcirclepos}\end{subfigure}~~~~~~~
\begin{subfigure}[b]{0.45\linewidth}~~~~~~~~~~~~~~\includegraphics[width=\textwidth]{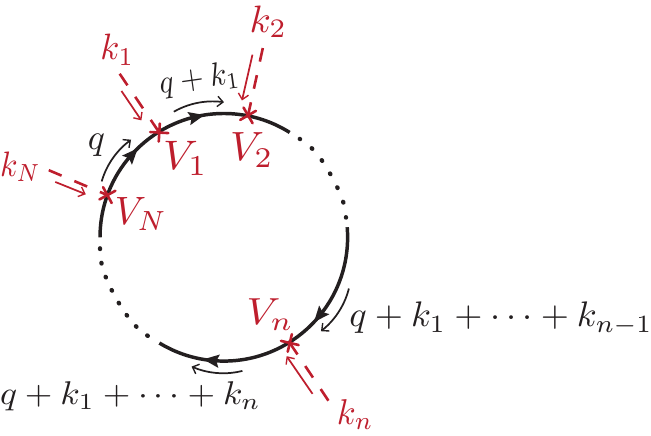}
\caption{In momentum space.}\label{fig:nonrenormcirclemom}\end{subfigure}~~~~~~~~~~~~
    \caption{Illustration of a subdiagram consisting of a single scalar loop.}\label{fig:nonrenorm}
\end{figure}

Consider a Feynman diagram that contains a {\it subdiagram} $\Gamma$ consisting of a scalar loop with the charge flowing along the loop, as illustrated in figure \ref{fig:nonrenormcirclepos}. We introduce insertions of $N$ local vertices along the loop labeled $V_n$ with $n=1 \,, \cdots\,, N$\,.
This scalar loop contributes the following factor to a Feynman diagram containing it,
\begin{equation} \label{eq:Gamma}
	\Gamma \propto \int \prod_{n=1}^N \, dt_n \, d^d \mathbf{x}_n \, V^{(n)} (t_n, \mathbf{x}_n) \, G (t_n- t_{n+1}, \mathbf{x}_n-\mathbf{x}_{n+1} )\,,
\end{equation}
where it is understood that $(t_{N+1}, \mathbf{x}_{N+1} ) \equiv (t_1, \mathbf{x}_1 )$ and the proportionality sign (rather than equality) is there to account for possible symmetry factors. Using the propagator \eqref{eq:posspaceprop} we obtain
\begin{align} \label{eq:thetaintegral}
	\Gamma &\propto \int \prod_{n=1}^N \, dt_n \, d^d \mathbf{x}_n \, \theta (t_n - t_{n+1} ) V^{(n)} (t_n, \mathbf{x}_n) \, G_W (t_n-t_{n+1}, \mathbf{x}_n-\mathbf{x}_{n+1})\,.
\end{align}
The above integral vanishes for $N \geq 2$ as long as the vertices do not contain time derivatives acting on the $\sigma$ and $\overline{\sigma}$ factors running in the loop. This is because the support of the integrand above is restricted to be at $t_1 = \ldots = t_N$, due to the $\theta$ functions. This support has measure zero in the domain of integration for $N\neq1$. On the other hand, when the vertex factors contain time derivatives, one can perform integrations by parts such that the time derivatives act on the $\theta$ functions, turning them into $\delta$ functions. In practice, each time derivative essentially eliminates an integral and so the integrals above vanish as long as the vertices (all together) contain no more than $N-2$ time derivatives acting on the $\sigma$ and $\overline{\sigma}$ factors running in the loop.

It is instructive to re-examine the same argument in momentum space. An illustration of the relevant scalar loop subdiagram $\Gamma$ appears in figure \ref{fig:nonrenormcirclemom} with the relevant momenta indicated. This subdiagram in a Feynman diagram contributes the following factor:
\begin{align} \label{eq:thetaintegralmom}
\begin{split}
\Gamma \propto 	
   \int \frac{d\nu}{2\pi} \,& \frac{d^d \mathbf{q}}{(2 \pi)^d} \, V^{(1)} (q, -q-k_1) \, \frac{i}{\nu+\omega_1 - \frac{|\mathbf{q} + \bk_1|^2}{2\Omega}+i \epsilon} \, \\
	&  \,  \times V^{(2)} (q+k_1, -q-k_1-k_2) \, \times \ldots \times V^{(N)} (q-k_N, -q) \, \frac{i}{\nu - \frac{|\mathbf{q}|^2}{2\Omega}+i \epsilon}\,,
\end{split}
\end{align}
where we have used the explicit form of the propagator in momentum space \eqref{propiepsilon} and labeled the vertices by the (incoming) momenta on the scalar lines.
Note that all the poles are located in the lower half $\nu$-plane. For $N \geq 2$\,, if the power of $\nu$ coming from the vertices in the numerator of the integrand is no higher than $N-2$\,, then at large $\nu$ the integrand decreases at least as fast as $|\nu|^{-2}$. We can then close the contour in the upper half-plane, and the integral in \eqref{eq:thetaintegralmom} evaluates to zero.

The vanishing of oriented scalar loops discussed above has far-reaching implications on the RG flow of sGED defined in eq.~\eqref{eq:sGEDaction}. First, note that sGED does not have any self-interactions of the gauge fields such as those in eq.~\eqref{eq:GEDint0}. In addition, the interaction vertices do not contain frequency insertions, thanks to our field redefinition in eq.~\eqref{eq:redef}. This immediately implies that the interactions in eq.~\eqref{eq:GEDint0} are not generated along the RG flow due to the fact that these corrections would be associated with 1PI diagrams in which all external legs are gauge fields. Such diagrams necessarily contain at least one scalar loop of the form in figure \ref{fig:nonrenorm}, which  evaluates to zero. The same logic also implies that there is no wavefunction renormalization for the gauge fields. The familiar argument based on gauge invariance fixes the beta function of the gauge coupling in terms of the wavefunction renormalization. This then implies that the gauge coupling $e$ does not renormalize. These conclusions hold at any loop order.

%% file: sections/RG_SGED.tex
In this section we study the one-loop quantum corrections in the theory of scalar Galilean electrodynamics \eqref{eq:sGEDaction}. In particular, our goal is to evaluate the beta functions for the infinitely-many couplings encoded in $\mathcal{J}[M]$, $\mathcal{V}[M]$ and $\mathcal{E}[M]$. We use a background-field method, expanding the Lagrangian order by order in the field $M$ -- or equivalently $\vp$ -- around a constant value. To perform this expansion, we treat the covariantized gap $M\equiv \Omega - e \varphi$ as a field with a large classical background value and a small fluctuation of the order of the perturbative coupling $e$, \ie we define
\begin{equation}
M = M_0 +\delta M\,,~~~ M_0=O(e^0)\,,~~~\delta M = O(e)~.
\end{equation}
In terms of the $\varphi$ field
\begin{align}\label{eq:defdeltaphi}
\begin{split}
M_0 & = \Omega - e \varphi_0~,\\
\delta M & = - e \delta \varphi~.
\end{split}
\end{align}
Note that a constant $\varphi = \varphi_0$ (together with all other fields vanishing) is a solution of the equations of motion \eqref{eq:sGEDeomat} for any choice of the functionals. This method allows us to express the beta functions of $\mathcal{J}[M]$, $\mathcal{V}[M]$ and $\mathcal{E}[M]$ in terms of these functionals and their derivatives.

\subsection{Feynman Rules}
Plugging \eqref{eq:defdeltaphi} in the action \eqref{eq:sGEDaction} and expanding around the background, we find the quadratic terms
\begin{align}
\begin{split}
S_{\text{free}}=&\,\int dt d^2 x \left(\frac{1}{2}(\del_t\delta \varphi)^2+E^i\partial_i \delta\varphi-\frac{1}{4}f^{ij}f_{ij}-\frac{1}{2\xi}(\partial_t\delta \varphi+\partial_i a_i)^2\right.
\\
&\,\left. +\frac{i}{2}\left(\bar \sigma \del_t \sigma-\sigma \del_t\bar \sigma\right)-\frac{1}{2M_0} \del_i\bar \sigma \del_i \sigma\right),
\end{split}
\end{align}
where we have also included a Galilean-invariant gauge-fixing term.
In the following, it will be convenient to organize the gauge fields in a vector
\begin{equation}
\mathcal{A}_I\equiv (\delta\varphi,a_t, a_i), \qquad i \in \{1,2\}\,,
\end{equation}
whose indices are denoted by capital letters $I, J, K, \cdots \in \{ \varphi, t, 1,2 \}$.
The resulting propagators are (with $k\equiv (\omega, {\bf k})$)
\begin{align}\label{propagators123}
\begin{split}
&\raisebox{-0.3\height}{\includegraphics[width=2.8in]{figures/ScalarPropagator.pdf}}
 \equiv \frac{i}{\omega - \frac{{\bf k}^2}{2M_0}+i\epsilon}~,
\\
&\includegraphics[width=3in]{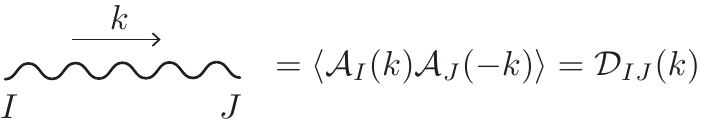}\\
&~~~~~~~~~~~~~~~~~~~~~~~~~~~~\equiv
-\frac{i}{{\bf k}^2}\left[
\begin{pmatrix}
0&1&0\\
1&0&0\\
0&0&\delta_{ij}
\end{pmatrix}
-\frac{(1-\xi)}{{\bf k}^2}
\begin{pmatrix}
0&0&0\\
0&\omega^2&-\omega {\bf k}_j\\
0&-\omega {\bf k}_i&{\bf k}_i {\bf k}_j
\end{pmatrix}\right]~.
\end{split}
\end{align}

The interaction Lagrangian is given by
\begin{align}
\begin{split}\label{eq:intlag}
\mathscr{L}_{\rm int} &= e a_t \bar{\sigma}\sigma - \left(\frac{1}{2 M}-\frac{1}{2 M_0}\right)\partial_i\bar{\sigma}\partial^i\sigma - \frac{1}{2M}\left[i e a_i(\bar{\sigma}\partial^i \sigma - \sigma\partial^i\bar{\sigma}) + e^2 a_i a^i \bar{\sigma}\sigma\right]\\
& ~~~~~~~~~~~~~+ \mathcal{J}[M]\partial_i M \partial^i M \,\bar{\sigma}\sigma -\frac{\lambda}{4} \mathcal{V}[M]\,(\bar{\sigma}\sigma)^2-\mathcal{E}[M](\del_i \del^i M - e^2 \bar \sigma \sigma)\, \overline{\sigma} \sigma~.
\end{split}
\end{align}
As we will see below, the coefficient $(2M)^{-1}$ in front of the spatial part of the kinetic term does not run, which is a consequence of boost invariance.\footnote{On the other hand one could also consider a rescaling of the full kinetic term as in eq.~\eqref{eq:totalaction} but this can be reabsorbed in a field redefinition as in \eqref{eq:redef} without changing the analysis of quantum corrections, as we show in appendix \ref{app:sGEDCM}.}
On the other hand, the coupling functions $\mathcal{J}[M]$ and $\mathcal{V}[M]$ are unspecified functions of $M$, and we will see that they run. We included in eq.~\eqref{eq:intlag} an additional operator that vanishes on-shell using the equation of motion \eqref{eq:eomvarphi} and whose coupling $\mathcal{E}[M]$ is also an arbitrary function of $M$. Even though such an operator does not contribute to any physical amplitude, it is important to include it to compute correctly the running of the physical couplings $\mathcal{J}[M]$ and $\mathcal{V}[M]$ as we will see below.

The expansion of eq.~\eqref{eq:intlag} in $\delta\varphi$ gives rise to infinitely many vertices, that are all classically marginal, but since each additional power of $\delta\varphi$ comes with an additional power of the coupling $e$, for a calculation at any finite order in $e$ only a finite number of vertices is needed. In particular, for the purpose of deriving the RG equations at leading order in perturbation theory, we only need vertices with at most five external legs. The relevant vertices with two scalar external legs are given by  (with $k_{1,2} \equiv (\omega_{1,2},{\bf k}_{1,2})$ and similarly for $p_{1,2}$)
\begin{align}
& \raisebox{-0.14\height}{\includegraphics[width=2in]{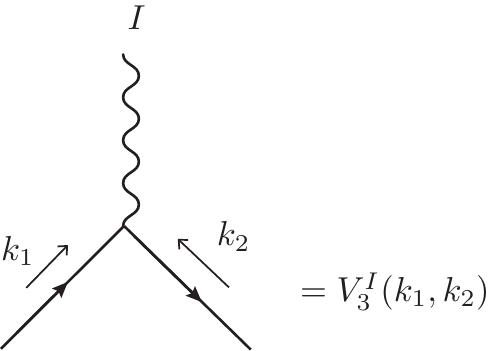}}~,\label{eq:cubicV} \\
&
\raisebox{-0.2\height}{\includegraphics[width=3.5in]{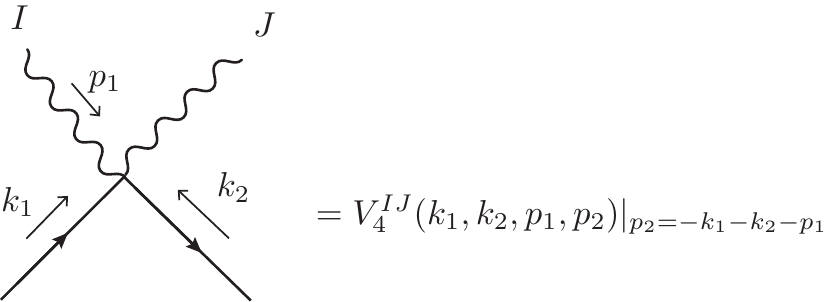}}~,\\
&\raisebox{-0.2\height}{\includegraphics[width=4in]{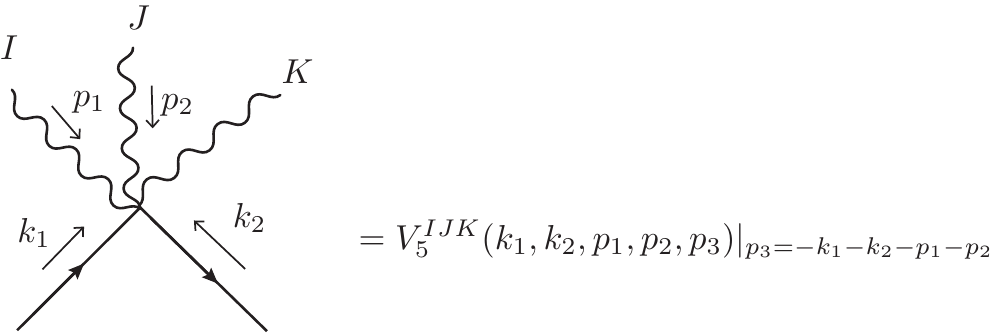}}~,
\end{align}
where
\begin{align}
\begin{split}
& V^I_3(k_1,k_2)  = i e
\begin{pmatrix}
\frac{1}{2 M_0^2} {\bf k}_1\cdot{\bf k}_2- \mathcal{E}(M_0) ({\bf k}_1+{\bf k}_2)^2\\1\\ \frac{1}{2 M_0}({\bf k}_1-{\bf k}_2)^i
\end{pmatrix}~,
\end{split}\\\notag\\
\begin{split}
& {V_4}^{IJ}(k_1,k_2,p_1,p_2)  = \frac{i e^2}{M_0}
\begin{pmatrix}
\frac{{\bf k}_1 \cdot {\bf k}_2 }{M_0^2} &0 &\frac{({\bf k}_1 - {\bf k}_2)_j}{2 M_0} \\
0 &0&0 \\
\frac{({\bf k}_1 - {\bf k}_2)_i}{2 M_0}&0&-\delta_{ij}
\end{pmatrix}
\\
&~~~~~~~~~~~~~~~~~~~~~~~~~~~~~~~~~~~
 +i e^2\left(\mathcal{E}'(M_0)({\bf p}_1^2+{\bf p}_2^2 ) - 2 \mathcal{J}(M_0) {\bf p}_1\cdot{\bf p}_2 \right)\delta^{I\varphi}\delta^{J\varphi}~,
\end{split}\label{eq:V4}\\\notag\\
\begin{split}
& {V_5}^{IJK} \equiv\frac{1}{3} \left(V_5^{IJ}\delta^{K\varphi} +V_5^{JK}\delta^{I\varphi} +V_5^{KI}\delta^{J\varphi} \right),
 \\
& V_5^{IJ}(k_1,k_2, p_1,p_2,p_3)  =  \frac{3i e^3}{M_0^2}\begin{pmatrix}
\frac{{\bf k}_1 \cdot {\bf k}_2 }{M_0^2} & 0 &\frac{({\bf k}_1 - {\bf k}_2)_j}{2 M_0} \\
0&0&0 \\
\frac{({\bf k}_1 - {\bf k}_2)_i}{2 M_0}&0&-\delta_{ij}
\end{pmatrix}\\
&~~~~~~
+ i e^3 \left(-\mathcal{E}''(M_0)({\bf p}_1^2+{\bf p}_2^2 +{\bf p}_3^2) + 2 \mathcal{J}'(M_0) \left({\bf p}_1\cdot{\bf p}_2+{\bf p}_2\cdot{\bf p}_3+{\bf p}_3\cdot{\bf p}_1 \right)\right)\delta^{I\varphi}\delta^{J\varphi}~.
\end{split}\label{eq:V5}
\end{align}
We will also need the following vertices with four scalar external legs
\begin{align}
& \raisebox{-0.39\height}{\includegraphics[width=1.5in]{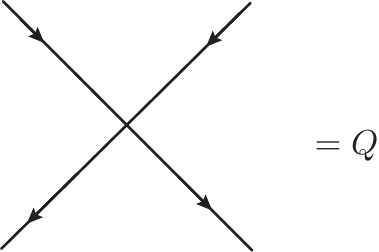}}~, \\
&\raisebox{-0.3\height}{\includegraphics[width=1.5in]{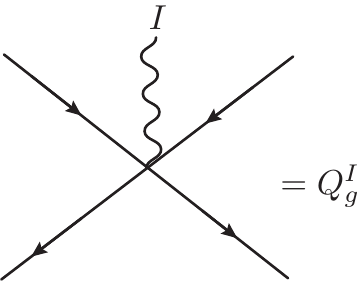}}~,
\end{align}
where
\begin{equation}\label{forZiqi}
Q  = - i \lambda \,\mathcal{V}(M_0)+ 4 i e^2\,\mathcal{E}(M_0)~,
\end{equation}
and the only non-zero entry of $Q_g^I$ is
\begin{equation}
Q_g^\varphi =  i \lambda \, e \,\mathcal{V}'(M_0)- 4 i e^3\,\mathcal{E}'(M_0)~.
\end{equation}

\subsection{Renormalization Constants}\label{renormconsts1}
Having obtained the Feynman rules, we can now proceed to derive the RG equations at leading order. To this end, we evaluate 1PI correlation functions, expand them in external momenta, and evaluate the left-over scale-independent integrals with a sharp UV cutoff $\Lambda$ and a sharp IR cutoff $\mu$. Further details on the regularization of the integrals can be found in appendix \ref{app:Integrals}.
In writing the correlation functions, we use ellipses to denote terms that we are neglecting because their order in the expansion in external momenta does not match the tree level result. Some of these terms include IR divergences that need to cancel in any physical observable, such as correlation functions of composite gauge-invariant operators.\footnote{Note that 1PI correlation functions of $\sigma$ and $\bar \sigma$ are not physical observables, because the operator $\sigma$ is not gauge invariant, and in fact, we will see that the result depends explicitly on the gauge-fixing parameter $\xi$.}

For the present purpose of renormalizing the 1PI correlation functions, we only retain the coefficient of the $\log\!\left(\Lambda\right)$ UV divergence and ignore IR divergent terms. In most cases the UV and IR divergences are neatly separated and canceling the $\log\!\left(\Lambda\right)$ is also sufficient to make the amplitude IR finite. However, as we will see, there is an exception in the four-point function of $\sigma$, where we also find a power-law IR and UV divergent contribution that we ignore. This exception is only present if the scalar field has bosonic statistics, because the quartic interaction vanishes in the case of fermionic statistics, while the renormalization of the other couplings is identical to the bosonic case, see the discussion in section \ref{sec:discussion}.

As this procedure of ignoring IR divergent terms might seem a bit ad-hoc, let us comment on it further.
Famously, in relativistic QED there are IR divergences which are usually resolved by the inclusion of soft-photons. However, in our case there are no asymptotic states associated to the gauge fields, and so it is unclear how to resolve the IR divergences. Recall however that the theory of GED can be obtained both from a non-relativistic limit with the addition of an extra real scalar or from a null reduction of four-dimensional Maxwell theory coupled to matter. One might then hope that the resolution of the IR problem in the parent theory, through the proper inclusion of soft photons, can be used to shed light on the IR divergences in GED. A similar problem is often discussed when studying scattering in non-relativistic effective field theory of QED, called NRQED in the literature, see e.g., \cite{Caswell:1985ui,Labelle:1996en}. There, the resolution is obtained by a matching procedure with the relativistic theory. However, at the first few orders in perturbation theory a correct result is obtained by simply ignoring the IR divergences, similarly to what we do here. It is possible that also in GED a complete definition of the theory requires matching conditions with the ``UV theory'' from which sGED can be obtained through the non-relativistic limit.\footnote{Similar divergences are also encountered in the quantization of gauge theories in the Coulomb gauge, and to solve this issue several authors have invoked the use of split dimensional regularization \cite{Leibbrandt:1996np,Leibbrandt:1997kh} in which both the dimensions of the temporal and the spatial sub-manifolds are analytically continued. This regularization is often used also in the context of non-relativistic theories \cite{Anselmi:2007ri,Arav:2016akx}. With the use of this technique also the divergence we find in the four-point function of $\sigma$ would be cured.}
We plan to examine this problem in more detail in the future.

\begin{figure}[tbp]
\center
\includegraphics[width=1.8in]{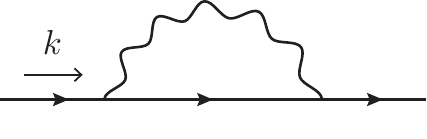}
    \caption{Corrections to the scalar propagator.}\label{fig:GED_diagrams_sigma_sigmabar0}
\end{figure}

The integral expressions and final results for each individual diagram used in this section are listed in appendix \ref{app:diagramsGED}.
Here, we only present the result for each 1PI correlation function, and for the corresponding renormalization constants.
To start with, we compute the one-loop correction to the propagator of the scalar, \ie the one-loop 1PI two-point function for the scalar field, given by the diagram in figure \ref{fig:GED_diagrams_sigma_sigmabar0}.\footnote{Here and in the following, various diagrams with gauge propagators starting and ending at the same vertex do not contribute to the calculation.}
We obtain
\begin{equation}
\begin{split}
\langle \sigma_B(k) \bar{\sigma}_B(-k)\rangle^{(1)}_{\text{1PI}}  = \frac{e^2}{2\pi}(1-\xi) J_0\, D_{\sigma}(k)^{-1}+\dots\,,
\end{split}
\label{eq:s2pt1loop}
\end{equation}
where the superscript $(1)$ indicates that this is the one-loop contribution, $J_0$ is the power-law divergent integral (see appendix \ref{app:Integrals} for more details)
\begin{equation}
J_0 \equiv \int \frac{d\nu}{2\pi} \int \frac{d|\mathbf{q}|}{|\mathbf{q}|^3}\,,
\end{equation}
and the subscript $B$ is used to denote the bare fields and couplings. Note that since the gauge-coupling does not run, as we have seen in section \ref{sec:nonrenorm}, we do not need to define a bare and a renormalized $e$, so we do not use any subscript on $e$. Similarly the gauge fields do not receive a wavefunction renormalization and therefore below we will not use the subscript $B$ for the gauge fields. Even though, as we just explained, we will only retain the $\log\!\left(\Lambda\right)$ divergence of the correlation functions, it is still useful to subtract the power-law divergence in \eqref{eq:s2pt1loop} with the wavefunction renormalization
\begin{equation}
\sigma_B = \sqrt{Z_\sigma} \, \sigma~,~~ \bar{\sigma}_B = \sqrt{Z_\sigma} \, \bar{\sigma}~,\label{eq:wfren}
\end{equation}
because this will allow us to check the cancellation of the gauge parameter $\xi$ from the remaining correlation functions, and because it will also cancel most of the contributions proportional to $J_0$, with the exception mentioned above, as we will see. Expanding $Z_\sigma = 1 + \delta Z_\sigma$, we find that
\begin{equation}
\delta Z_\sigma = \frac{e^2}{2\pi}(1-\xi) J_0~,\label{eq:dZSGED}
\end{equation}
cancels the divergence in \eqref{eq:s2pt1loop}.

\begin{figure}[tbp]
\center
\hspace{1cm}\begin{subfigure}[bc]{0.2\linewidth}\includegraphics[width=\linewidth]{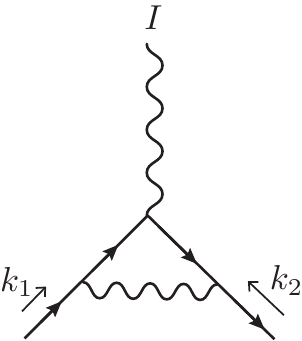}\caption{}\label{fig:sged_diagrams_3pta}
\end{subfigure}\hspace{20mm}
\begin{subfigure}[bc]{0.2\linewidth}\includegraphics[width=\linewidth]{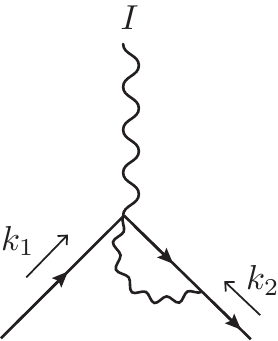}\caption{}\label{fig:sged_diagrams_3ptb}
\end{subfigure}\hspace{20mm}
\caption{Corrections to the three-point function.}
\label{fig:sged_diagrams_3pt}
\end{figure}

Next, we consider the one-loop correction to the cubic vertex \eqref{eq:cubicV}, \ie the one-loop 1PI three-point function of one gauge field and two scalars, given by the diagrams in figure \ref{fig:sged_diagrams_3pt}.
We obtain
\begin{align}
\begin{split}
&\langle \sigma_B(k_1) \bar{\sigma}_B(k_2) \mathcal{A}^I(-k_1-k_2)\rangle^{(1)}_{\text{1PI}} = -\frac{e^2 }{2\pi}(1-\xi) J_0\, V_3^I(k_1,k_2)\\
& ~~~~~~~~~~~~~~~~~~~~~~~~~+ i\frac{e^3 }{2\pi} \delta^{I\vp} ({\bf k}_1+{\bf k}_2)^2 \left( \mathcal{E}'_B[M_0]+\frac{1}{4 M_0^3}\right) \log \left(\frac{\Lambda }{\mu }\right)+\dots~.
\end{split}\label{eq:1loop3pt}
\end{align}
We define the renormalized coupling $\mathcal{E}[M]$ by
\begin{equation}\label{deltaE}
-\mathcal{E}_B[M](\del_i \del^i M - e^2 \bar \sigma_B \sigma_B)\, \overline{\sigma}_B \sigma_B = -(\mathcal{E}[M]+\delta\mathcal{E}[M])(\del_i \del^i M - e^2\, Z_\sigma\, \bar \sigma \sigma)\, Z_\sigma\, \overline{\sigma} \sigma\,.
\end{equation}
Requiring the cancellation of the UV divergences in eq.~\eqref{eq:1loop3pt}, we fix the counterterm to be
\begin{equation}
\delta\mathcal{E}[M] = \frac{e^2}{2\pi} \left( \mathcal{E}'[M] +\frac{1}{4 M^3}\right) \log \left(\frac{\Lambda }{\mu }\right)~.\label{eq:deltaI}
\end{equation}
Note that all the $\xi$-dependent divergences in eq.~\eqref{eq:1loop3pt} are subtracted by the renormalization of the scalar external legs, and as a result the counterterm \eqref{eq:deltaI} for the vertex is $\xi$-independent. This is expected because $\delta \mathcal{E}[M]$ determines the running of a gauge-invariant (set of) coupling(s), and it provides a nice consistency check of the calculation.

\begin{figure}
\centering
\begin{subfigure}[b]{0.27\linewidth}\includegraphics[width=\linewidth]{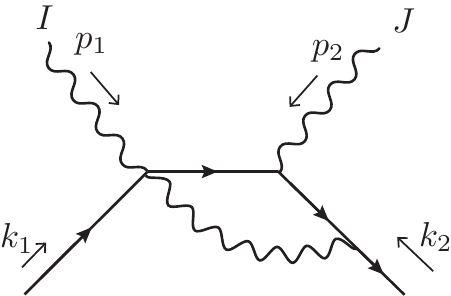}\caption{}\label{fig:sged_2sigma_2gaugea}\end{subfigure}
~~~
\begin{subfigure}[b]{0.27\linewidth}\includegraphics[width=\linewidth]{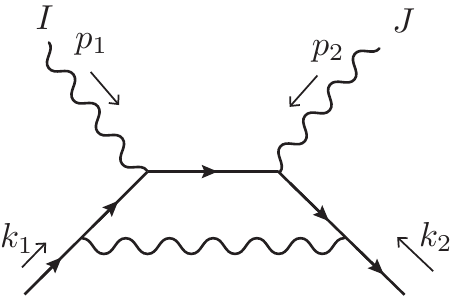}\caption{}\label{fig:sged_2sigma_2gaugeb}
\end{subfigure}
~~~
\begin{subfigure}[b]{0.25\linewidth}\includegraphics[width=\linewidth]{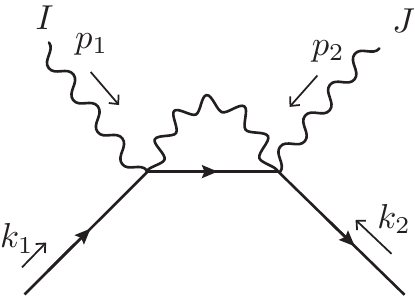}\caption{}\label{fig:sged_2sigma_2gaugec}\end{subfigure}
\\
\center
\begin{subfigure}[b]{0.22\linewidth}\includegraphics[width=\linewidth]{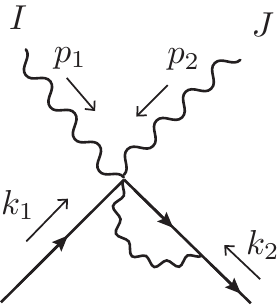}\caption{}
\label{fig:sged_2sigma_2gauged}
\end{subfigure}~~~~~
\begin{subfigure}[b]{0.22\linewidth}\includegraphics[width=\linewidth]{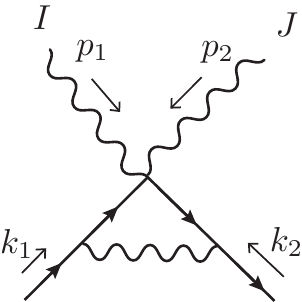}\caption{}
\label{fig:sged_2sigma_2gaugee}
\end{subfigure}
    \caption{Corrections to the four-point function of two scalars and two gauge fields.}\label{fig:sged_2sigma_2gauge}
\end{figure}

Next, we consider the one-loop corrections to the 1PI four-point function of two scalars and two gauge fields. The Feynman diagrams are shown in figure \ref{fig:sged_2sigma_2gauge}.
Summing up the diagrams we find
\begin{align}
\begin{split}
 \langle \sigma_B(k_1) \bar{\sigma}_B(k_2) &\mathcal{A}^I(p_1) \mathcal{A}^J(p_2)\rangle^{(1)}_{\text{1PI}} = -\frac{e^2 }{2\pi}(1-\xi) J_0\, V_4^{IJ}(k_1,k_2 ,p_1,p_2)
\\ &+ i\frac{e^4}{2\pi}\,\delta^{I\varphi}\delta^{J\varphi}\left[-({\bf p}^2_1+{\bf p}^2_2)\left(\mathcal{E}_B''[M_0]-\frac{3}{4 M_0^4}\right)\right.\\&~~~~~~~~~~~~~~~~~~~~~~\left.+2\,{\bf p}_1\cdot {\bf p}_2\, \left(\mathcal{J}_B'[M_0]+\frac{5}{8 M_0^4}\right) \right]\log \left(\frac{\Lambda }{\mu }\right)+\dots~,
\end{split}\label{eq:1loop4pt}
\end{align}
where a momentum-conserving delta function is implicit, \ie $p_2=-k_1-k_2-p_1$.
Comparing this expression to the $\varphi\varphi$ entry of the quartic vertex in eq.~\eqref{eq:V4}, we observe that  the correction \eqref{eq:deltaI} to the function $\mathcal{E}[M]$ is precisely what is needed to cancel the UV divergence proportional to $({\bf p}^2_1+{\bf p}^2_2)$ in eq.~\eqref{eq:1loop4pt}.
We proceed by defining the renormalized coupling $\mathcal{J}[M]$ in terms of the bare quantities as follows
\begin{equation}
\mathcal{J}_B[M]\partial_i M \partial^i M \,\bar{\sigma}_B\sigma_B = Z_\sigma(\mathcal{J}[M]+\delta\mathcal{J}[M])\partial_i M \partial^i M\,\bar{\sigma}\sigma~.
\end{equation}
Finally, requiring the cancellation of the UV divergence proportional to ${\bf p}_1\cdot {\bf p}_2$ we obtain
\begin{equation}
\delta\mathcal{J}[M] = \frac{e^2}{2\pi} \left(\mathcal{J}'[M] +\frac{5}{8 M^4}\right) \log \left(\frac{\Lambda }{\mu }\right)~.
\end{equation}
As above, all the $\xi$ dependence has canceled using the wavefunction renormalization of the $\sigma$ field.

\begin{figure}
\centering
\begin{subfigure}[b]{0.25\linewidth}\includegraphics[width=\linewidth]{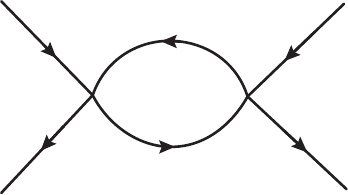}\caption{}\label{fig:sigma4ptSCSa}
\end{subfigure}~~~~
\begin{subfigure}[b]{0.1\linewidth}\includegraphics[width=\linewidth]{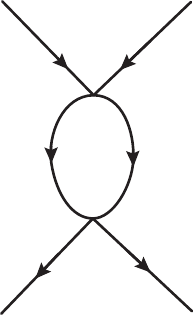}\caption{}\label{fig:sigma4ptSCSb}
\end{subfigure}~~~~
\begin{subfigure}[b]{0.17\linewidth}\includegraphics[width=\linewidth]{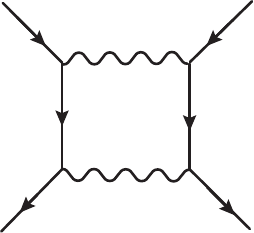}\caption{}\label{fig:sigma4ptSCSc}
\end{subfigure}~~~~
\begin{subfigure}[b]{0.17\linewidth}\includegraphics[width=\linewidth]{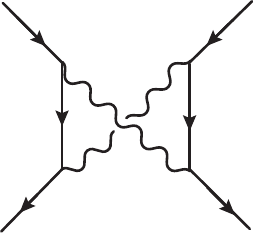}\caption{}\label{fig:sigma4ptSCSd}
\end{subfigure}~~~~
\begin{subfigure}[b]{0.17\linewidth}\includegraphics[width=\linewidth]{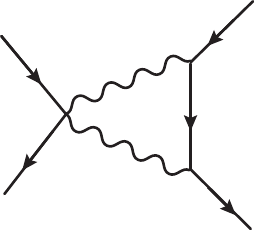}\caption{}\label{fig:sigma4ptSCSe}
\end{subfigure}
\\\vspace{10pt}
\begin{subfigure}[b]{0.25\linewidth}\includegraphics[width=\linewidth]{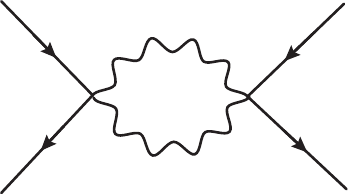}
\caption{}\label{fig:sigma4ptSCSf}
\end{subfigure}~~~~~
\begin{subfigure}[b]{0.15\linewidth}\includegraphics[width=\linewidth]{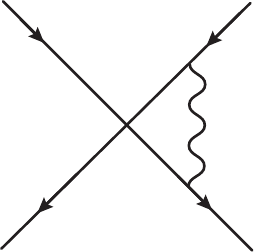}\caption{}\label{fig:sigma4ptSCSg}
\end{subfigure}~~~~~
\begin{subfigure}[b]{0.15\linewidth}\includegraphics[width=\linewidth]{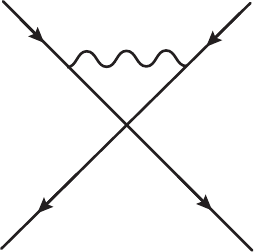}\caption{}\label{fig:sigma4ptSCSh}
\end{subfigure}~~~~~
\begin{subfigure}[b]{0.15\linewidth}\includegraphics[width=\linewidth]{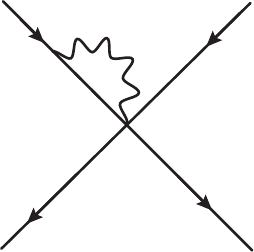}\caption{}\label{fig:sigma4ptSCSi}
\end{subfigure}~~~~~
\caption{Corrections to the four-point function of $\sigma$.}
\label{fig:sigma4pt}
\end{figure}

Finally, we consider the 1PI four-point function of four scalar fields required for the renormalization of the $(\bar{\sigma}\sigma)^2$ coupling. The corresponding diagrams are shown in figure \ref{fig:sigma4pt}. The result is
\begin{align}
\begin{split}
&\langle\sigma_B\bar{\sigma}_B\sigma_B\bar{\sigma}_B\rangle^{(1)}_{\text{1PI}} \\ & = \left[\frac{i e^4}{2\pi}\left( - 4\mathcal{E}'_B[M_0]+4 \mathcal{J}_B[M_0]-\frac{3}{8  M_0^3}\right)  +  \frac{i \lambda e^2}{2\pi} \left(2\mathcal{V}'_B[M_0]+\frac{\mathcal{V}_B[M_0]}{2 M_0}\right) \right. \\ & ~~~~~~~~\left. +\frac{i \lambda^2}{4\pi} M_0 \mathcal{V}_B[M_0]^2 \right]\log \left(\frac{\Lambda }{\mu }\right)-   \frac{ e^2}{\pi}  (1-\xi) J_0   Q +\frac{i e^4}{\pi}\frac{ J_0 }{M_0^2}+\dots~.
\end{split}\label{eq:sigma4pt}
\end{align}
The renormalized quartic coupling $\mathcal{V}[M]$ is defined by
\begin{equation}
 -\frac{\lambda}{4}\mathcal{V}_B[M]\,(\bar{\sigma}_B\sigma_B)^2 =  -\frac{\lambda}{4}  Z_\sigma^2 \left(\mathcal{V}[M]+ \delta\mathcal{V}[M]\right)(\bar{\sigma}\sigma)^2~.
\end{equation}
Once again, the $\xi$ dependence in \eqref{eq:sigma4pt} completely cancels with the wavefunction renormalization \eqref{eq:dZSGED}. As mentioned earlier, we see that there is also a leftover power-law UV and IR divergent term proportional to $J_0$ in the last line of \eqref{eq:sigma4pt}. Here we will only reabsorb the more physical $\log\!\left(\Lambda\right)$ divergence in the renormalization of the quartic coupling for the purpose of computing its beta function as explained at the beginning of the section.
Requiring the cancellation of the logarithmic UV divergence and using the counter term for $\delta \mathcal{E}$ in eq.~\eqref{eq:deltaI}, we obtain
\begin{align}
\hspace{-7pt}\lambda\delta\mathcal{V}[M] =  \left[\frac{e^4}{2\pi}\left( 4 \mathcal{J}[M] + \frac{5}{8 M^3}\right) +   \frac{\lambda e^2}{2\pi} \left(2 \mathcal{V}'[M]+\frac{\mathcal{V}[M]}{2  M}\right)+\frac{ \lambda ^2}{4\pi} M \mathcal{V}[M]^2 \right]\log \left(\frac{\Lambda }{\mu }\right).
\end{align}

\subsection{RG Equations and Fixed Points}\label{ref:Shira}

From the renormalization constants computed in the previous section we obtain the beta functions\footnote{In our conventions, the beta function of a coupling $g$ is defined to be $\beta_g =\frac{d \,\delta g}{d\log\Lambda}$.}
\begin{align}
\begin{split}
\beta_{\mathcal{J}[M]}&=\frac{e^2}{2\pi}\left(\mathcal{J}'[M]+\frac{5}{8 M^4}\right)~,\\
\beta_{\lambda \mathcal{V}[M]}& =\frac{e^4}{2\pi}\left( 4 \mathcal{J}[M] + \frac{5}{8 M^3}\right) +   \frac{\lambda e^2}{2\pi} \left(2 \mathcal{V}'[M]+\frac{\mathcal{V}[M]}{2  M}\right)+\frac{ \lambda ^2}{4\pi} M \mathcal{V}[M]^2 ~.
\end{split}\label{eq:bSGED}
\end{align}
Similarly for the EOM-vanishing operator we have
\begin{equation}
\beta_{\mathcal{E}[M]}=\frac{e^2}{2\pi}\left(\mathcal{E}'[M] +\frac{1}{4 M^3}\right) ~.\label{eq:bEOM}
\end{equation}
Note that on general grounds the EOM-vanishing coupling should not enter the beta function of the physical couplings (see, \eg the explanation around eqs.~(6.40)-(6.41) in \cite{Manohar:2018aog}), and our calculation indeed confirms this. Besides these couplings there is also the gauge coupling $e$, that according to the general arguments of section \ref{sec:nonrenorm} has a vanishing beta function.

We now proceed to look for fixed points, \ie zeroes of the full set of beta functions. The coupling $e$ is a free parameter because its beta function vanishes identically. We need to keep this free parameter small in order for the one-loop calculation of the beta functions of the other couplings to be reliable. In other words, we can only explore the existence of fixed points in the perturbative region $e\ll 1$ of the parameter space. Setting the beta functions of the couplings $\mathcal{J}$ and $\lambda\mathcal{V}$ to zero leads to differential equations for $\mathcal{J}[M]$ and $\lambda\mathcal{V}[M]$. Rather than writing down the most general solution to these differential equations, we observe that a simple set of solutions is obtained by taking $\mathcal{J}$ and $\mathcal{V}$ to be proportional to powers of $M$:
\begin{equation}\label{fp1}
\mathcal{J}[M] = j \,M^{-3}~,~~\mathcal{V}[M] = v \,M^{-2}~,
\end{equation}
where $j$ and $v$ are real constants (\ie they are $M$-independent). Setting $\beta_{\mathcal{J}[M]} = 0$ simply fixes
\begin{equation}\label{fp3}
j=j_* \equiv \frac{5}{24}~.
\end{equation}
Plugging this solution and the ansatz for $\mathcal{V}$ in $\beta_{\lambda \mathcal{V}[M]}=0$, the whole $M$-dependence becomes just an overall factor of $M^{-3}$ and we obtain an algebraic equation for the parameter $v$,
\begin{equation}
12\,\lambda^2 v^2 - 84 \,e^2 \lambda v +  35 \,e^4 = 0~,
\end{equation}
with solutions
\begin{equation}\label{fp2}
\lambda v^\pm_* \equiv \frac{21 \pm 4 \sqrt{21}}{6} e^2~.
\end{equation}
Note that for both choices of sign $\lambda v$ is positive, ensuring that the quartic potential for $\sigma$ is stable.
Similarly, a simple fixed-point solution for the EOM-vanishing operator is found using the power-law ansatz $\mathcal{E}[M] = \varepsilon\, M^{-2}$. Plugging this ansatz in the beta function \eqref{eq:bEOM} and requiring it to vanish fixes $\varepsilon =\varepsilon_*\equiv \frac 18$.

More generally, we could allow for arbitrary integration constants when solving the equations for the fixed point, \eg $\mathcal{J}[M] = j \,M^{-3} + \mathcal{J}_0$ and similarly for $\mathcal{E}[M]$ and $\mathcal{V}[M]$ (though the form of the general solution for $\mathcal{V}[M]$ is more cumbersome due to the nonlinearity of the equation). It might be useful to explain the origin of these parameters from the point of view of the infinite series of couplings. Upon expanding $\mathcal{J}[M]=\sum_{n=0}^\infty \frac{(M-\Omega)^n}{n!}\mathcal{J}_n$, the integration constant $\mathcal{J}_0$ corresponds to the coupling with the lowest power of $M-\Omega = -e\varphi$. Expressing the beta function of $\mathcal{J}[M]$ in terms of the component couplings, we see that the beta  function of each $\mathcal{J}_n$ depends only on $\mathcal{J}_{n+1}$, and that $\mathcal{J}_0$ does not enter any beta function. Therefore,  the coupling $\mathcal{J}_{0}$ remains as a free parameter at the fixed point even after setting all the beta functions $\beta_{\mathcal{J}_i}=0$. Including these integration constants in addition to the gauge coupling $e^2$ gives us a four-parameter space of exactly marginal couplings parameterizing a non-relativistic conformal manifold.

The solutions \eqref{fp1}-\eqref{fp2} for $\mathcal{V}[M]$ and $\mathcal{J}[M]$ were obtained by setting to zero the beta functions at the lowest order in an expansion in the gauge-coupling $e$, so the position of the fixed point will receive corrections at higher order in $e$.
It can sometimes happen that beta functions accidentally vanish identically at one-loop and this gives rise to an ``approximate'' manifold of fixed-points, which then disappears at the next loop order. However, we would like to emphasize that the origin of the manifold of fixed-points in sGED is different since none of the beta functions in eqs.~\eqref{eq:bSGED}-\eqref{eq:bEOM} vanish at one loop. Rather, we found non-trivial solutions for all the beta function equations. As a consequence, the manifold of fixed-points persists to higher loop orders and the location of the fixed points can be systematically corrected order-by-order in perturbation theory in the small parameter $e$.

It is interesting to examine the RG stability of the fixed points \eqref{fp1}-\eqref{fp2}, \ie whether the classically marginal couplings (other than those parameterizing the conformal manifold) are marginally relevant or marginally irrelevant at the non-trivial fixed points. With finitely many couplings, the RG stability is determined by the signs of the eigenvalues of the matrix of derivatives of the beta functions $\sim \partial_i \beta^j$, with $i,j$ indices that run over the set of classically marginal couplings. Hence the natural adaptation to our setting is to consider functional variations of the beta functionals in eq.~\eqref{eq:bSGED}: we vary the beta functionals with respect to the coupling functions, and look at the signs of the eigenvalues of the operator obtained via this functional variation.
Plugging $\mathcal{J}[M] = j_* M^{-3} + \delta \mathcal{J}[M]$ and $\mathcal{V}[M] = v^{\pm}_* M^{-2}+ \delta \mathcal{V}[M]$ in eq.~\eqref{eq:bSGED}, and expanding to linear order in the variation, we obtain the following eigenvalue equations:
\begin{equation}
\begin{split}
\delta \beta_{\mathcal{J}[M]} &\equiv \frac{e^2}{2\pi}\delta \mathcal{J}' = \alpha \delta \mathcal{J}\,,\\
\delta \beta_{\lambda \mathcal{V}[M]} &\equiv \frac{2 e^4}{\pi} \delta \mathcal{J}+ \frac{e^2}{\pi} \lambda\delta \mathcal{V}' + \frac{\lambda}{4\pi M}(e^2 + 2 \lambda v^{\pm}_*)  \delta \mathcal{V}
=\alpha \lambda \delta \mathcal{V}\,,
\end{split}
\end{equation}
from which we would like to determine the possible signs of the eigenvalues $\alpha$. Unfortunately, unlike the finite dimensional case, this question cannot be answered unless we supplement the equations with boundary conditions, or equivalently unless we somehow specify the space of functions on which the operators act. It is not clear to us how to determine these additional conditions from first principles. A possible requirement is that the variations do not blow up for large covariantized mass $M$.
With these boundary conditions, there is no solution for $\alpha>0$, while for any fixed value of $\alpha<0$, there is a two-dimensional space of eigenfunctions, meaning that there are two relevant (functional) deformations of the non-trivial fixed point.\footnote{By contrast, around the free theory $e^2 = 0, \lambda = 0$ the coupling $\mathcal{J}[M]$ is exactly marginal, because its beta function vanishes identically for $e^2 = 0$, while the quartic coupling $\mathcal{V}[M]$ is marginally irrelevant because the only term in the beta function for $e^2=0$ is $\beta_{\lambda \mathcal{V}[M]}\vert_{e^2 = 0} = \frac{\lambda^2}{4\pi}M \mathcal{V}[M]^2$,  which is positive definite.} The case $\alpha=0$ corresponds to moving inside the manifold of fixed points and it is special because for this value of $\alpha$ we only find a one-dimensional space of eigenfunctions consistent with our boundary conditions. Therefore, contrary to the expectation that the two integration constants give two independent directions on the conformal manifold, postulating the above boundary condition implies that motion along one of these directions is not allowed.

%% file: sections/discussion.tex
In this paper we explored the quantum properties of Galilean electrodynamics \cite{Santos:2004pq,Festuccia} coupled to a Schr\"odinger scalar in 2+1 dimensions.
The theory consists of temporal and spatial gauge fields as well as an additional scalar $\vp$. This scalar $\vp$ is dimensionless (in a $z=2$ anisotropic sense) and completely  inert under all the symmetries of the problem.
It turns out that the theory generates an infinite series of quantum corrections proportional to different powers of the field $\vp$.

It is interesting to examine these corrections from the perspective of null reduction. Recall that the theory of Galilean electrodynamics can be obtained by considering a relativistic Maxwell gauge field in $3+1$ dimensions where the fields are assumed to be independent of the null coordinate $x^+$ and the other null coordinate $x^-$ takes the role of time in the non-relativistic setup. The field $\vp$ is simply the $A_+$ component of the higher dimensional gauge field. Despite being disallowed in the higher dimensional parent theory, here couplings proportional to $\vp^n$ are no longer forbidden by the gauge symmetry in the $x^+$ direction.

We focused on a subset of the possible quantum corrections, represented by the Lagrangian in eq.~\eqref{eq:sGEDaction}, where the couplings were packed into three functions $\mathcal{J}[M]$, $\mathcal{V}[M]$ and  $\mathcal{E}[M]$ (recall $M\equiv \Omega-e \vp$), with different terms in the Taylor expansion around $M=\Omega$ capturing interactions proportional to different powers of the field $\vp$.
We observed that the electromagnetic coupling $e$ does not run due to the non-renormalization theorem explained in section \ref{sec:nonrenorm}. This result holds to all loop orders.
We explained how to renormalize the theory using diagrammatic techniques, by systematically expanding the coupling functions around a fixed background value $\vp=\vp_0$ in such a way that the vertices and propagators depend on $\vp_0$. In this way, every Feynman diagram represents infinitely many  diagrams which arise when further expanding the vertices in terms of the background value.

Our analysis resulted in explicit expressions for the beta functions of the couplings $\mathcal{J}[M]$, $\mathcal{V}[M]$ and $\mathcal{E}[M]$ at one loop order, which we expressed in terms of the functions themselves and their derivatives, see eqs.~\eqref{eq:bSGED}-\eqref{eq:bEOM}.
The beta functions can be set to zero in order to find fixed-points. The special fixed points \eqref{fp1}-\eqref{fp2} correspond to a stable quartic potential $\mathcal{V}[M](\bar \sigma \sigma)^2$. The general solutions are characterized by integration constants thus providing a four-parameter family of fixed points (including the electromagnetic coupling $e$).\footnote{See, however, the comments at the end of section \ref{ref:Shira}.}
As explained in the introduction, the appearance of conformal manifolds is rare and here it is associated with the non-relativistic nature of the theory.

Regarding possible applications of the theory, we note that the presence of the dimensionless field $\vp$ implies that the fixed points have infinitely many relevant deformations, \eg $\vp^n \bar \sigma \sigma$, which would need to be fine-tuned in order to reach criticality.\footnote{This includes the $n=0$ term, which represents a chemical potential. It would be interesting to explore the influence of the chemical potential (and its exotic cousins $n\neq 0$) on the dynamics of the system.} It is therefore unlikely that this theory in its current form will describe quantum critical points in real-world condensed matter systems.
Various natural extensions of sGED can be formulated, including generalizations to non-abelian gauge groups, \ie Galilean Yang-Mills theory along the lines of \cite{Bagchi:2015qcw,Gomis:2020fui}, supersymmetric extensions \cite{SUSYGED}, and multi-flavored versions, with either bosonic or fermionic statistics.
It is an interesting direction for the future to see if one can eliminate the unwanted relevant deformations in any of these extensions. In this respect, we view our work as a first step in the exploration of the landscape of Schr\"odinger-invariant QFTs. Moreover, the equations of motion for GED naturally emerge when studying non-relativistic string theory,\footnote{See \cite{Gomis:2000bd, Danielsson:2000gi, Danielsson:2000mu} for the original works on non-relativistic string theory as well as \cite{Bergshoeff:2019pij} and references therein for a review of recent progress.}
where the extra scalar mode $\varphi$ appears as the Nambu-Goldstone boson associated with the spontaneous breaking of the translational symmetry in the coordinate transverse to a D-brane \cite{Gomis:2020fui}.

Due to the absence of a spin-statistics theorem in non-relativistic theories, a Schr\"odinger scalar could also have fermionic statistics. Therefore it is interesting to ask how our analysis is modified if we take fermionic statistics for the scalar coupled to GED. The calculations of the one-loop corrections follow closely those in this paper, with the following changes: the operator $(\bar \sigma \sigma)^2$ now vanishes so we cannot write down the coupling(s) $\mathcal{V}[M]$, and we need to include additional minus signs due to the fermionic statistics when summing over permutations of the external legs (due to the non-renormalization theorem, we never encounter closed fermionic loops). The result of these modifications is that the beta function of $\mathcal{V}[M]$ vanishes automatically as expected, because the associated diagrams do not depend on external momenta and therefore they vanish when anti-symmetrizing in the external legs, while the other beta functions are not modified. Note that the fermionic quartic coupling would not vanish identically if we allow for $N_f >1$ copies of the matter fields, namely in the theory with flavor symmetry, and this case deserves a separate analysis that we leave for the future.

One might worry that some of the couplings that parametrize the manifold of fixed points can be absorbed in a field redefinition, in which case they would be unphysical. In order to exclude this possibility, we need to see how they enter in physical observables. As an example, we can consider the 2-to-2 scattering amplitude of $\sigma$ particles in sGED. This analysis follows closely the one of \cite{Bergman} for the theory of a Schr\"odinger scalar coupled to the Chern-Simons term in $2+1$ dimensions. At tree level, working in the center of mass frame we obtain
\begin{equation}
i \mathcal{M}_{\mathrm{tree}}=-i \lambda \mathcal{V}\left[M_{0}\right]-\frac{i e^{2}}{2 M_{0}^{2}} \cdot \frac{1+3 \cos ^{2} \theta}{\sin ^{2} \theta}\,.
\end{equation}
Note that only the charge-to-mass ratio $e/M_0$ appears in the part of the amplitude that depends on the scattering angle $\theta$, while  the 4-point coupling $\lambda  \mathcal{V}[M_0]$ only appears in the constant piece. Therefore, the 4-point coupling $\lambda  \mathcal{V}[M_0]$ and the charge-to-mass ratio $e/M_0$ are bona-fide observable parameters and their physical values can be extracted from a 2-to-2 scattering
experiment. It would be interesting to further study the scattering-problem at one-loop (in which case the coupling $\mathcal{J}[M]$ should come into play), and understand the cancellation of IR divergences along the lines discussed in section \ref{renormconsts1}.

Finally, an interesting task for the future is to clarify the nature of the non-relativistic quantum mechanics associated to the sGED quantum field theory. In order to study this problem, one needs to derive the Schr\"odinger equation for a certain fixed number of $\sigma$ particles and in particular determine the form of their potential, induced by the interactions mediated by the GED gauge fields. Note that the gauge fields couple precisely to the $U(1)$ current associated to the number conservation symmetry, implying that $N>0$ states always include a flux for the gauge fields at infinity. It would be interesting to clarify how this affects the map to the quantum-mechanical problem.

%% file: sections/appGens.tex
The currents for the Schr\"odinger symmetry can be written in terms of the energy-momentum tensor and $U(1)$ current as follows, see \eg \cite{Nakayama:2013is}:
\begin{subequations}
\begin{align}
    \mathcal{\MN}^{t} &= - J_{m}^{t}, &%
    \mathcal{\MN}^{i} &= - J_{m}^{i}, \\
    \mathcal{H}^{t} &= T^{t}{}_{t}, &%
    \mathcal{H}^{i} &= T^{i}{}_{t}, \\
    \mathcal{P}^{t}{}_{i} &= T^{t}{}_{i}, &%
    \mathcal{P}^{j}{}_{i} &= T^{j}{}_{i}, \\
    \mathcal{J}^{t}{}_{ij} &= - x_i T^{t}{}_{j} + x_j T^{t}{}_{i}, &%
    \mathcal{J}^{k}{}_{ij} &= - x_i T^{k}{}_{j} + x_j T^{k}{}_{i}, \\
    \mathcal{G}^{t}{}_{i} &= - t \, T^{t}{}_{i} + x_i J_m^t, &%
    \mathcal{G}^{j}{}_{i} &= - t \, T^{j}{}_{i} + x_i T^{tj}, \label{Ggen}\\
    \mathcal{D}^{t} &= 2 t \, T^{t}{}_{t} + x^{i} T^{t}{}_{i}, &%
    \mathcal{D}^{i} &= 2 t \, T^{i}{}_{t} + x^j T^{i}{}_{j}, \\
    \mathcal{C}^{t} &= t^2 T^{t}{}_{t} + tx^i T^{t}{}_{i} - \frac{1}{2} x^2 J_m^t, &%
    \mathcal{C}^{i} &= t^2 T^{i}{}_{t} + tx^j T^{i}{}_{j} - \frac{1}{2} x^2 T^{ti}.
\end{align}
\end{subequations}
Using the conservation of the energy-momentum tensor \eqref{eq:consT} and $U(1)$ current \eqref{eq:Jconservation}, as well as the identity \eqref{eq:TJWard}, the trace condition \eqref{eq:Ttrace}, and the symmetry of the spatial components of the energy-momentum tensor, $T_{ij} = T_{ji}$, one can show that each one of the above currents is conserved. It is understood that these currents are evaluated using the energy momentum tensor which has been improved to satisfy the trace condition $2 T^t{}_t + T^i{}_i = 0$\,. The generators themselves are simply the charges associated with these currents:
\begin{equation}
    \bigl\{ \MN, H, P_i , J_{ij}, G_i , D, C \bigr\} = \int d^d x \, \bigl\{ \mathcal{\MN}^{t}, \mathcal{H}^t , \mathcal{P}^t, \mathcal{J}^{t}{}_{ij}, \mathcal{G}^{t}{}_{i} , \mathcal{D}^t , \mathcal{C}^t \bigr\}.
\end{equation}
These generators form the Schr\"{o}dinger algebra in eq.~\eqref{eq:Schrodinger}.

%% file: sections/appIntegrals.tex
In this appendix, we collect results for the integrals used in section \ref{sec:quantumGED}, which can always be brought to the form
\be\label{basicInt}
	I_{n,m} \equiv \int \frac{d\nu}{2\pi} \frac{d^2 \mathbf{q}}{(2\pi)^2} \, \frac{\nu^{2n-m-2} \, |\mathbf{q}|^{2m}}{(\nu^2 + |\mathbf{q}|^4)^n}\,,
		\qquad
	n = 0\,, 1\,, 2\,, \ldots\,
	\quad	\text{and} \quad
	m \in \mathbb{Z}\,,
\ee
where the variable $\nu$ is the frequency, Wick rotated to Euclidean signature. The Wick rotation is consistent with the $+i\epsilon$ prescription which we used to define our propagator \eqref{propagators123}. In order to obtain the form \eqref{basicInt} we have symmetrized the integrands in $\nu\to-\nu$. This is allowed given that we are using a regularization that preserves the symmetry $\nu\to-\nu$.
This also implies that when $m$ is odd this integral vanishes identically. When $m = 2 m'$ is even, we write
\be
	\nu^{2(n-m'-1)} = \ls (\nu^2 + |\mathbf{q}|^4) - |\mathbf{q}|^4 \rs^{n-m'-1}\,,
\ee
and then take a binomial expansion.
It follows that
\be \label{eq:Inm}
	I_{n,2m'} = \frac{1}{2\pi} \sum_{\ell=0}^{n-m'-1} (-1)^\ell \frac{(n-m'-1)!}{(n-m'-\ell-1)! \, \ell!} \, J_{m'+\ell+1}\,,
\ee
where we have defined
\be \label{eq:Jn}
	J_{n} \equiv 2 \pi \int \frac{d\nu}{2\pi} \, \frac{d^2 \mathbf{q}}{(2\pi)^2} \, \frac{|\mathbf{q}|^{4(n-1)}}{(\nu^2 + |\mathbf{q}|^4)^n}\,,
		\qquad
	n = 0\,, 1\,, 2\,, \ldots\,.
\ee

Next, let us evaluate the integrals \eqref{eq:Jn}. When $n=0$\,, we find
\be \label{eq:CJ0}
	J_0 = \int \frac{d\nu}{2\pi} \int \frac{d|\mathbf{q}|}{|\mathbf{q}|^3}\,.
\ee
The integral $J_0$ is superficially log-divergent by an anisotropic $z=2$ power-counting. However, due to the singular behavior of the
momentum pole in eq.~\eqref{eq:CJ0}, which does not depend on the frequency, $J_0$ does not contain any log divergences. Instead, it depends crucially on how one regulates it in the UV, which suggests that it should not contain any universal information about beta functions. For example, using the sharp cutoff regularization ($\mu \leq |\mathbf{q}| \leq \Lambda$ and $0 \leq a^2|\nu| \leq \Lambda^2$, where $a$ simply parametrizes the possible discrepancy in the temporal and spatial cutoffs), one finds a power law divergence in $J_0$ that depends on the ratio of UV and IR cutoffs.
As we explain below, for $n \geq 1$\,, we find
\be \label{eq:CJn}
	J_n = \frac{1}{2\sqrt{\pi}} \frac{\Gamma(n-\tfrac{1}{2})}{\Gamma(n)} \log \! \lr \frac{\Lambda}{\mu} \rr + \text{finite}\,,
\ee
where $\Lambda$ is a UV momentum cutoff and $\mu$ is an IR regulator. Note that in the limit $n \rightarrow 0$\,, the log divergence in eq.~\eqref{eq:CJn} becomes zero,
which is indeed consistent with eq.~\eqref{eq:CJ0} not being log divergent. Substituting eq.~\eqref{eq:CJn} into eq.~\eqref{eq:Inm}, we find
\be
	I_{n,m} = \frac{1}{4\pi^2} \frac{\Gamma(\frac{m+1}{2}) \, \Gamma( n - \frac{m+1}{2} )}{\Gamma(n)} \log \! \lr \frac{\Lambda}{\mu}\rr + \text{finite},
		\qquad%
	n = 1, 2, \ldots, \quad
		\text{and $m$ even.}
\ee

Now, we return to the derivation of the log divergence \eqref{eq:CJn} using a number of different sharp cutoff regularizations. Unlike for relativistic quantum field theories, where one often chooses a spherical regularization in Euclidean frequency-momentum space, here we can select different shapes of the cutoff surface in frequency-momentum space, all invariant under spatial rotations.
It was argued in \cite{Berthier:2017slt}, that the log divergence in the integrals $J_n$ above is independent of the shape of the cutoff surface for a certain class of ``star-shaped'' cutoff surfaces (in frequency-momentum space).\footnote{Though \cite{Berthier:2017slt} focuses on Lifshitz fixed points with $z=3$, this claim is valid for any value of the critical exponent $z$.} In particular, this claim was proven in appendix A of \cite{Berthier:2017slt} for the case of $n=2$. In the following, we apply the same procedure to study the integrals $J_n$ with $n \geq 1$\,, and show that the result in \eqref{eq:CJn} is independent of the shape of the cutoff for a large family of cutoff surfaces. We start by considering a simple oval-shaped cutoff surface as a warm-up, see ``method 1'' below. Then, we discuss the proof for bounded ``star-shaped'' cutoff surfaces, see ``method 2''.  Finally, we consider another natural (unbounded) choice of cutoff, see ``methods 3'' below.

\vspace{3mm}

\noindent {\bf Method 1:} We first consider the following cutoff surface
\begin{align}
	\mathcal{V}_1 = \left\{ (|\nu|, |\mathbf{q}|): \quad \mu^4 \leq a^4 \nu^2 + |\mathbf{q}|^4 \leq \Lambda^4 \right\}\,,
\end{align}
and take the change of variables from $(|\nu|\,, |\mathbf{q}|)$ to $(r\,, \theta)$\,,
\begin{align}\label{changeOfvarsappB}
	|\nu| = \frac{\Lambda^2}{a^2} \, r \sin \theta\,,
		\qquad%
	|\mathbf{q}|^2 = \Lambda^2 \, r \cos \theta\,,
\end{align}
where $\Lambda$ plays the role of a UV cutoff, $\mu$ plays the role of an IR cutoff and $a$ is an arbitrary constant fixing the ovality of the cutoff surface in frequency-momentum space. Note that this shape of the cutoff surface nicely fits with anisotropic $z=2$ dimensional analysis with $[\Lambda]=[\mu]=1$.
In terms of the variables \eqref{changeOfvarsappB}, the domain of integration becomes
\begin{equation}
    \mathcal{V}_1 = \left\{ (r,\theta): \quad 0 \leq \theta \leq \pi/2, \qquad \mu^2 / \Lambda^2 \leq r \leq 1\right\}\,,
\end{equation}
and the integrals  $J_n$ in eq.~\eqref{eq:Jn} can be rewritten as\footnote{Note that here the $|\nu|$ integral only runs over the positive real axis.}
\begin{align}\label{integralchangeB2}
	J_n & =\frac{1}{\pi} \int_{\mathcal{V}_1} d|\nu| \, d|\mathbf{q}| \, \frac{|\mathbf{q}|^{4n-3}}{(\nu^2 + |\mathbf{q}|^4)^{n}}
 =\frac{1}{2 \pi} \, \int_{\mathcal{V}_1}dr \, d\theta \, \frac{a^{4n-2}}{r} \frac{(\cos\theta)^{2(n-1)}}{\lr a^4 \cos^2 \theta + \sin^2 \theta \rr^{\! n}}\,.
\end{align}
Performing the $\theta$ and $r$ integrals explicitly yields,
\begin{align}
\begin{split}
	J_n
	& = \frac{1}{2\sqrt{\pi}} \, \frac{\Gamma\lr n-\frac{1}{2} \rr}{\Gamma \lr n \rr} \log \! \lr\frac{\Lambda}{\mu} \rr.
\end{split}
\end{align}

\vspace{3mm}

\noindent \textbf{Method 2:} Now we follow closely the procedure outlined in \cite{Berthier:2017slt} and study a more general class of cutoff surfaces, generalizing the surface considered above in ``method 1''.
As in \cite{Berthier:2017slt}, we regularize the integral in the star-shaped domain
\begin{equation}
    \mathcal{V}_2 = \left\{ (r,\theta): \quad 0 \leq \theta \leq \pi/2, \qquad g(\theta) \,\mu^2 / \Lambda^2 \leq r \leq f(\theta) \right\}\,,
\end{equation}
given in terms of the variables \eqref{changeOfvarsappB}, where $f(\theta)$ and $g(\theta)$ are two arbitrary functions parameterizing the shape of the cutoff surface in the UV and IR, respectively. We will further assume that these functions are order one, positive and bounded and satisfy $\mu \, g(\theta)< \Lambda f(\theta)$ for all values of $\theta$.\footnote{We assume that these functions are order one so that the IR cutoff remains small and the UV cutoff remains large. More formally, we assume that $f(\theta)$ and $g(\theta)$ are such that the integral over the term that contains $\log \! \lr {f(\theta)}/{g(\theta)}\rr$ in eq.~\eqref{fg123} gives a finite contribution.}
The functions define the shape of the cutoff surface in the positive quadrant of the integration plane $0 \leq \theta \leq \pi/2$.
The integrals $J_n$ in eq.~\eqref{integralchangeB2} become
\begin{align}
\begin{split}\label{fg123}
	J_n & = \frac{a^{4n-2}}{2 \pi} \, \int_0^\frac{\pi}{2}  \, \frac{(\cos\theta)^{2(n-1)}d\theta}{\lr a^2 \cos^2 \theta + \sin^2 \theta \rr^{\! n}} \int_{g(\theta)\frac{\mu^2}{\Lambda^2}}^{ f(\theta)} \frac{dr}{r}
\\
	& = \frac{a^{4n-2}}{2 \pi} \, \int_0^\frac{\pi}{2}  \, \frac{(\cos\theta)^{2(n-1)}d\theta}{\lr a^2 \cos^2 \theta + \sin^2 \theta \rr^{\! n}}
\left[ 2\log \! \lr \frac{\Lambda}{\mu} \rr
+ \log \! \lr \frac{f(\theta)}{g(\theta)} \rr\right]	
\\ & = \frac{1}{2\sqrt{\pi}} \frac{\Gamma(n-\frac{1}{2})}{\Gamma(n)} \log \! \lr \frac{\Lambda}{\mu} \rr +\text{finite}\,.
\end{split}
\end{align}
Note that $f(\theta)$ and $g(\theta)$ influence only the scheme-dependent finite part of the result. Hence, we demonstrated that the coefficient of the log divergence in eq.~\eqref{eq:CJn} is independent of the detailed choice of the cutoff surface.

\vspace{3mm}

\noindent {\bf Method 3:} So far, we only considered compact integration domains. Next, we consider an unbounded integration domain in frequencies given by
\be
	\mathcal{V}_3 = \Bigl \{ (|\nu|, |\mathbf{q}|) : \quad 0 \leq |\nu| < \infty\,, \quad \mu \leq |\mathbf{q}| \leq \Lambda \Bigr \}\,,
\ee
in which case we obtain once again
\begin{align}
	J_n &
= \frac{1}{\pi} \int_\mu^\Lambda d|\mathbf{q}| \int_{0}^\infty d|\nu| \, \frac{|\mathbf{q}|^{4n-3}}{(\nu^2 + |\mathbf{q}|^4)^{n}} 
		 = \frac{1}{2\sqrt{\pi}} \, \frac{\Gamma\lr n-\frac{1}{2}\rr}{\Gamma(n)} \log \! \lr \frac{\Lambda}{\mu} \rr.
\end{align}

%% file: sections/diagramsGED.tex
In this appendix we list the explicit results for the one-loop diagrams that are used in section \ref{sec:quantumGED} (see figures \ref{fig:GED_diagrams_sigma_sigmabar0}-\ref{fig:sigma4pt}) to compute the $\beta$ functions in sGED.
It should be understood that each figure represents the sum of the set of diagrams obtained by permutations of the identical external legs and/or by charge conjugation.

\begin{itemize}[label={}]
\item{ {\bf$\boldsymbol{\langle \sigma (k) \, \overline{\sigma} (-k) \rangle}$\,:}
{\small{\begin{equation}
\!\!\!\!\!\!\!\!\!\!\!\!\!\!\!\!\!\!\!\!\!\!\!\!\!\!\!\text{Fig. \ref{fig:GED_diagrams_sigma_sigmabar0}} = \frac{e^2}{2\pi}J_0 (1-\xi )(\mathcal{D}_\sigma(k))^{-1}~.
\end{equation}}}}
\item{{\bf$\boldsymbol{\langle \sigma (k_1) \, \overline{\sigma} (k_2) \mathcal{A}^I(-k_1-k_2) \rangle}$\,:}
{\small{\begin{align}
\begin{split}
\text{Fig. \ref{fig:sged_diagrams_3pta}} & =- \frac{e^2(1-\xi )J_0}{2 \pi }  V_3^I(k_1,k_2)+\frac{i e^3(1-\xi )\delta^{I\varphi} \log \left(\frac{\Lambda }{\mu }\right)}{8 \pi M_0^2 } \left(\omega_1-\omega_2-\frac{{\bf k}_1^2+{\bf k}_2^2}{2M_0}\right)~,
\\
\text{Fig. \ref{fig:sged_diagrams_3ptb}} & = \frac{i e^3 \delta^{I\varphi} \log \left(\frac{\Lambda }{\mu }\right)}{2 \pi M_0^2}  \left[ M_0^2\left( \mathcal{E}'[M_0] +\frac{1}{4 M_0^3}\right)  ( {\bf k}_1+   {\bf k}_2)^2\right.\\&\hspace{6cm}\left.-\frac{(1-\xi)}{4 } \left(\omega_1-\omega_2-\frac{{\bf k}_1^2+{\bf k}_2^2}{2M_0}\right)\right]~.
\end{split}
\end{align}}}}
\item{{\bf$\boldsymbol{\langle \sigma (k_1) \, \overline{\sigma} (k_2) \mathcal{A}^I(p_1) \mathcal{A}^J(p_2 = -k_1-k_2-p_1) \rangle}$\,:}

{\small{
\begin{align}
\begin{split}
\text{Fig. \ref{fig:sged_2sigma_2gaugea}} & = -\frac{i e^4(1-\xi)\log \left(\frac{\Lambda }{\mu }\right)}{4 \pi M_0^2 }
\begin{pmatrix}
\frac{3 {\bf k}_1 \cdot {\bf k}_2 - {\bf p}_1 \cdot {\bf p}_2}{M_0^2} &1 &\frac{3({\bf k}_1 - {\bf k}_2)_j}{2 M_0} \\
1 &0&0 \\
\frac{3({\bf k}_1 - {\bf k}_2)_i}{2 M_0}&0&-2 \delta_{ij}
\end{pmatrix}\\&\hspace{2.5cm}  +\frac{i e^4(1-\xi)\log \left(\frac{\Lambda }{\mu }\right)}{4 \pi M_0^2 }\left(\mathcal{E}[M_0]+\frac{3}{4 M_0^2}\right)({\bf p}_1^2+{\bf p}_2^2)\delta^{I\varphi}\delta^{J\varphi}~,
\end{split}\nonumber\\\nonumber\\
\begin{split}
\text{Fig. \ref{fig:sged_2sigma_2gaugeb}}  & = \frac{i e^4(1-\xi)\log \left(\frac{\Lambda }{\mu }\right)}{4 \pi M_0^2 }
\begin{pmatrix}
\frac{2 {\bf k}_1 \cdot {\bf k}_2 - \frac 34 {\bf p}_1 \cdot {\bf p}_2}{M_0^2} &1 &\frac{({\bf k}_1 - {\bf k}_2)_j}{ M_0} \\
1 &0&0 \\
\frac{({\bf k}_1 - {\bf k}_2)_i}{M_0}&0&- \delta_{ij}
\end{pmatrix}\\&\hspace{2.5cm}  -\frac{i e^4(1-\xi)\log \left(\frac{\Lambda }{\mu }\right)}{4 \pi M_0^2 }\left(\mathcal{E}[M_0]+\frac{1}{2 M_0^2}\right)({\bf p}_1^2+{\bf p}_2^2)\delta^{I\varphi}\delta^{J\varphi}~,
\end{split}\nonumber\\\nonumber\\
\begin{split}
\text{Fig. \ref{fig:sged_2sigma_2gaugec}} & = -\frac{i e^4(1+\xi)\log \left(\frac{\Lambda }{\mu }\right)}{4 \pi M_0^2 }
\begin{pmatrix}
\frac{{\bf k}_1 \cdot {\bf k}_2 - \frac 14 ({\bf p}_1 \cdot {\bf p}_2+{\bf p}_1^2+{\bf p}_2^2)}{M_0^2} &0 &\frac{({\bf k}_1 - {\bf k}_2)_j}{ 2M_0} \\
0 &0&0 \\
\frac{({\bf k}_1 - {\bf k}_2)_i}{2M_0}&0&- \delta_{ij}
\end{pmatrix}~,
\end{split}\\\nonumber\\
\begin{split}
\text{Fig. \ref{fig:sged_2sigma_2gauged}} & = \frac{i e^4\log \left(\frac{\Lambda }{\mu }\right)}{4 \pi M_0^2 }\begin{pmatrix}
\frac{(1+\xi){\bf k}_1 \cdot {\bf k}_2-(1-\xi)M_0(\omega_1-\omega_2)}{M_0^2}&0&\frac{({\bf k}_1 - {\bf k}_2)_j}{M_0}\\
0 &0&0 \\
\frac{({\bf k}_1 - {\bf k}_2)_i}{M_0}&0&- 2\delta_{ij}
\end{pmatrix}\\&\hspace{-2cm} +\frac{i e^4\log \left(\frac{\Lambda }{\mu }\right)}{4 \pi }\left[\left(4\mathcal{J}'[M_0]+\frac{3-\xi}{M_0^4}\right){\bf p}_1 \cdot {\bf p}_2 -2\left(\mathcal{E}''[M_0]-\frac{3-\xi}{4 M_0^4}\right) ({\bf p}_1^2+{\bf p}_2^2)\right]\delta^{I\varphi}\delta^{J\varphi}~,
\end{split}\nonumber\\\nonumber\\
\begin{split}
\text{Fig. \ref{fig:sged_2sigma_2gaugee}} & = \frac{i e^4(1-\xi)\log \left(\frac{\Lambda }{\mu }\right)}{4 \pi M_0^2 }\begin{pmatrix}
\frac{{\bf k}_1 \cdot {\bf k}_2 - \frac 12 ({\bf p}_1 +{\bf p}_2)^2 + M_0(\omega_1-\omega_2)}{M_0^2}&0&0\\
0 &0&0 \\
0&0&0
\end{pmatrix}\\& \hspace{6cm}-\frac{e^2(1-\xi) J_0}{2\pi} V_4^{IJ}(k_1,k_2,p_1,p_2)~.
\end{split}\nonumber
\end{align}}}}
\item{{\bf$\boldsymbol{\langle \sigma  \, \overline{\sigma} \sigma \bar{\sigma}  \rangle}$\,:}
{\small{
\begin{align}
\begin{split}
\text{Fig. \ref{fig:sigma4ptSCSa}} &= 0~,
\end{split}\nonumber\\
\begin{split}
\text{Fig. \ref{fig:sigma4ptSCSb}} & = \frac{i  }{4\pi} \log \left(\frac{\Lambda }{\mu }\right)M_0(\lambda \mathcal{V}[M_0] - 4 e^2\mathcal{E}[M_0])^2~,
\end{split}\nonumber\\
\begin{split}
\text{Fig. \ref{fig:sigma4ptSCSc}} & = \frac{4i e^4}{\pi} M_0\left(\mathcal{E}[M_0]+\frac{1}{8 M_0^2}\right)\left[\log \left(\frac{\Lambda }{\mu }\right)\left(\mathcal{E}[M_0]+\frac{1}{8 M_0^2}\right)-\frac{(1-\xi)J_0 }{M_0}\right]~,
\end{split}\nonumber\\
\begin{split}
\text{Fig. \ref{fig:sigma4ptSCSd}} & = \frac{4i e^4(1-\xi)}{\pi} J_0\left(\mathcal{E}[M_0]-\frac{\xi}{8 M_0^2}\right) ~,
\end{split}\nonumber\\
\begin{split}
\text{Fig. \ref{fig:sigma4ptSCSe}} & = \frac{2i e^4}{\pi}\left[\log \left(\frac{\Lambda }{\mu }\right)\left(\mathcal{E}'[M_0]+\mathcal{J}[M_0]-\frac{1}{8 M_0^3}\right)+\frac{(1-\xi^2)J_0 }{2M_0^2}\right] ~,
\end{split}\\
\begin{split}
\text{Fig. \ref{fig:sigma4ptSCSf}} & = \frac{i e^4}{2\pi}\frac{(1+\xi^2)J_0}{M_0^2}~,
\end{split}\nonumber\\
\begin{split}
\text{Fig. \ref{fig:sigma4ptSCSg}} & = \frac{2i e^2}{\pi}(\lambda \mathcal{V}[M_0] - 4 e^2\mathcal{E}[M_0])(1-\xi)J_0 ~,
\end{split}\nonumber\\
\begin{split}
\text{Fig. \ref{fig:sigma4ptSCSh}} & = \frac{2i e^2}{\pi}M_0(\lambda \mathcal{V}[M_0] - 4 e^2\mathcal{E}[M_0])\left[\log \left(\frac{\Lambda }{\mu }\right)\left(\mathcal{E}[M_0]+\frac{1}{8 M_0^2}\right)-\frac{(1-\xi)J_0 }{2M_0}\right]~,
\end{split}\nonumber\\
\begin{split}
\text{Fig. \ref{fig:sigma4ptSCSi}} & = \frac{i e^2}{\pi}\log \left(\frac{\Lambda }{\mu }\right)(\lambda \mathcal{V}'[M_0] - 4 e^2\mathcal{E}'[M_0]) ~.
\end{split}\nonumber
\end{align}}}
}
\end{itemize}

%% file: sections/sGEDCM.tex
In section \ref{sec:GED}, we introduced the following field redefinition \eqref{eq:redef} of $\sigma$ (and $\overline{\sigma}$),
\be \label{eq:sigmafred}
	\sigma \rightarrow \frac{\sigma}{\sqrt{\mathcal{C[M]}}} \,,
\ee
which eliminates $\mathcal{C} [M]$ in the sGED action \eqref{eq:totalaction}. This procedure allowed us to focus on the running of the couplings $\mathcal{J}[M]$\,, $\lambda \CV [M]$ and $\mathcal{E} [M]$ in the action \eqref{eq:sGEDaction}. In this appendix, we revisit this field redefinition and calculate the running of $\mathcal{C} [M]$\,. Moreover, we will confirm that the running of $\mathcal{C}[M]$ does not affect the beta functions or the family of fixed points we found in section \ref{ref:Shira}, and thus further justify the classical field redefinition taken in eq.~\eqref{eq:sigmafred}.

Consider the action (with a gauge fixing term),\footnote{We chose the form of the $\mathcal{\tilde E}[M]$ term such that also for the action \eqref{eq:sGEDCM} it is proportional to an equation of motion operator obtained by varying the action with respect to $a_t$. This choice simplifies some of the calculations described below, but does not imply any loss of generality.}
\begin{align} \label{eq:sGEDCM}
\begin{split}
\hspace{-5pt} S_\text{sGED} = & \int dt \, d^2 \mathbf{x} \,  \biggl\{ \frac{1}{2} \dot\vp^2 + E^i \p_i \varphi - \frac{1}{4} f^{ij} f_{ij} -\frac{1}{2\xi}(\dot \varphi+\partial_i a_i)^2
 \\
 &+ \mathcal{C}[M] \! \ls \frac{i}{2} \bigl( \overline{\sigma} D_t \sigma - \sigma D_t \overline{\sigma} \bigr) - \frac{1}{2M} D_i \overline{\sigma} D^i \sigma \rs \\[2pt]
	& \! + \tilde{\mathcal{J}}[M] \, \p_i M \p^i M \, \overline{\sigma} \sigma - \frac{1}{4} \, \lambda \, \tilde{\mathcal{V}}[M] \, (\overline{\sigma} \sigma)^2 - \tilde{\mathcal{E}} [M] \bigl( \p_i \p^i M - e^2 \, \mathcal{C}[M]  \, \overline{\sigma} \sigma \bigr) \, \overline{\sigma} \sigma \biggr\},
\end{split}
\end{align}
which is related to the action \eqref{eq:sGEDaction} used in the bulk of the paper by the field redefinition \eqref{eq:sigmafred}, with the couplings $\tilde{\mathcal{J}}[M]$\,, $\lambda \, \tilde{\mathcal{V}}[M]$ and $\tilde{\mathcal{E}} [M]$ related to $\mathcal{J}[M]$\,, $\lambda \, \mathcal{V}[M]$ and $\mathcal{E} [M]$ in \eqref{eq:sGEDaction} according to
\begin{equation}
\begin{split}\label{eq:cbJVE}
	\mathcal{J}[M] & = \frac{\tilde{\mathcal{J}}[M]}{\mathcal{C}[M]} - \frac{1}{8 M} \left(\frac{\mathcal{C}'[M]}{\mathcal{C}[M]}\right)^{\!2}
-\frac{1}{4} \left(\frac{\mathcal{C}'[M]}{M \mathcal{C}[M]}\right)^{\!\prime}\,, \\
	\lambda {\CV}[M] & = \frac{\lambda \tilde{\mathcal{V}}[M]}{\mathcal{C}^2[M]} + \frac{e^2}{M} \frac{\mathcal{C}'[M]}{\mathcal{C}[M]}\,, \quad 
	\mathcal{E}[M]  = \frac{\tilde{\mathcal{E}}[M]}{\mathcal{C}[M]} + \frac{1}{4M} \frac{\mathcal{C}'[M]}{\mathcal{C}[M]}\,.
\end{split}
\end{equation}
Starting with the action \eqref{eq:sGEDCM} and following the same procedure presented in section \ref{sec:quantumGED}, we find that the one-loop beta function of $\mathcal{C}[M]$ is
\be
	\beta_{\mathcal{C}[M]} = \frac{e^2}{4 \pi} \, \mathcal{C}'[M]\,.
\ee
This result requires choosing the same wavefunction renormalization as in eq.~\eqref{eq:dZSGED}.\footnote{One may also set the running of $\mathcal{C}[M]$ to zero by introducing a wavefunction renormalization that is different from \eqref{eq:dZSGED},
\be
	\delta Z_\sigma = \frac{e^2}{2\pi} \ls (1-\xi) J_0 + \frac{\mathcal{C}'[M]}{2 \, \mathcal{C}[M]} \log \frac{\Lambda}{\mu} \rs\,,
\ee
which absorbs the running of $\mathcal{C}[M]$ completely. This is why the classical field redefinition \eqref{eq:cbJVE} is justified quantum mechanically. The freedom to divide the quantum corrections between $\delta Z_\sigma$ and $\mathcal{C}[M]$ is analogous to choosing a particular set of coordinates in the target space of nonlinear sigma models.}
The beta functions of  $\tilde{J}[M]$\,, $\lambda \tilde{V}[M]$ and $\tilde{\mathcal{E}}[M]$ can also be obtained by following the procedure outlined in section \ref{sec:quantumGED}. These expressions are a bit lengthy and we will not write them explicitly here. However, there is a significant simplification after applying the change of basis \eqref{eq:cbJVE} to change the variables from $\tilde{J}[M]$\,, $\lambda \tilde{V}[M]$ and $\tilde{\mathcal{E}}[M]$ to $\mathcal{J}[M]$\,, $\lambda \, \mathcal{V}[M]$ and $\mathcal{E} [M]$. As one may expect, it turns out that the beta functions for the couplings $\mathcal{J}[M]$\,, $\lambda \, \mathcal{V}[M]$ and $\mathcal{E} [M]$ are the same as those given in eqs.~\eqref{eq:bSGED} and \eqref{eq:bEOM}.
At the fixed point, where $\beta_{\mathcal{C}[M]} = 0$, we find that $C[M]$ is a constant. In this case the redefinition in eqs.~\eqref{eq:sigmafred} and \eqref{eq:cbJVE} becomes a constant rescaling of the Schr\"odinger field and couplings. Of course, after performing the rescaling, we obtain the family of fixed points which were discussed in section \ref{ref:Shira}.

%% file: RGGED.bbl
\providecommand{\href}[2]{#2}\begingroup\raggedright\begin{thebibliography}{10}

\bibitem{Poland:2018epd}
D.~Poland, S.~Rychkov and A.~Vichi, \emph{{The Conformal Bootstrap: Theory,
  Numerical Techniques, and Applications}},
  \href{https://doi.org/10.1103/RevModPhys.91.015002}{\emph{Rev. Mod. Phys.}
  {\bfseries 91} (2019) 015002}
  [\href{https://arxiv.org/abs/1805.04405}{{\ttfamily 1805.04405}}].

\bibitem{mcgreevy2010pursuit}
J.~McGreevy, \emph{In pursuit of a nameless metal}, {\emph{Physics} {\bfseries
  3} (2010) 83}.

\bibitem{Hartnoll:2009ns}
S.~A. Hartnoll, J.~Polchinski, E.~Silverstein and D.~Tong, \emph{{Towards
  strange metallic holography}},
  \href{https://doi.org/10.1007/JHEP04(2010)120}{\emph{JHEP} {\bfseries 04}
  (2010) 120} [\href{https://arxiv.org/abs/0912.1061}{{\ttfamily 0912.1061}}].

\bibitem{Hagen:1972pd}
C.~Hagen, \emph{{Scale and conformal transformations in galilean-covariant
  field theory}}, \href{https://doi.org/10.1103/PhysRevD.5.377}{\emph{Phys.
  Rev. D} {\bfseries 5} (1972) 377}.

\bibitem{Mehen:1999nd}
T.~Mehen, I.~W. Stewart and M.~B. Wise, \emph{{Conformal invariance for
  nonrelativistic field theory}},
  \href{https://doi.org/10.1016/S0370-2693(00)00006-X}{\emph{Phys. Lett.}
  {\bfseries B474} (2000) 145}
  [\href{https://arxiv.org/abs/hep-th/9910025}{{\ttfamily hep-th/9910025}}].

\bibitem{Nishida:2007pj}
Y.~Nishida and D.~T. Son, \emph{{Nonrelativistic conformal field theories}},
  \href{https://doi.org/10.1103/PhysRevD.76.086004}{\emph{Phys. Rev.}
  {\bfseries D76} (2007) 086004}
  [\href{https://arxiv.org/abs/0706.3746}{{\ttfamily 0706.3746}}].

\bibitem{Golkar:2014mwa}
S.~Golkar and D.~T. Son, \emph{{Operator Product Expansion and Conservation
  Laws in Non-Relativistic Conformal Field Theories}},
  \href{https://doi.org/10.1007/JHEP12(2014)063}{\emph{JHEP} {\bfseries 12}
  (2014) 063} [\href{https://arxiv.org/abs/1408.3629}{{\ttfamily 1408.3629}}].

\bibitem{Goldberger:2014hca}
W.~D. Goldberger, Z.~U. Khandker and S.~Prabhu, \emph{{OPE convergence in
  non-relativistic conformal field theories}},
  \href{https://doi.org/10.1007/JHEP12(2015)048}{\emph{JHEP} {\bfseries 12}
  (2015) 048} [\href{https://arxiv.org/abs/1412.8507}{{\ttfamily 1412.8507}}].

\bibitem{Pal:2018idc}
S.~Pal, \emph{{Unitarity and universality in nonrelativistic conformal field
  theory}}, \href{https://doi.org/10.1103/PhysRevD.97.105031}{\emph{Phys. Rev.}
  {\bfseries D97} (2018) 105031}
  [\href{https://arxiv.org/abs/1802.02262}{{\ttfamily 1802.02262}}].

\bibitem{Gubler:2015iva}
P.~Gubler, N.~Yamamoto, T.~Hatsuda and Y.~Nishida, \emph{{Single-particle
  spectral density of the unitary Fermi gas: Novel approach based on the
  operator product expansion, sum rules and the maximum entropy method}},
  \href{https://doi.org/10.1016/j.aop.2015.03.007}{\emph{Annals Phys.}
  {\bfseries 356} (2015) 467}
  [\href{https://arxiv.org/abs/1501.06053}{{\ttfamily 1501.06053}}].

\bibitem{SonWingate}
D.~T. Son and M.~Wingate, \emph{{General coordinate invariance and conformal
  invariance in nonrelativistic physics: Unitary Fermi gas}},
  \href{https://doi.org/10.1016/j.aop.2005.11.001}{\emph{Annals Phys.}
  {\bfseries 321} (2006) 197}
  [\href{https://arxiv.org/abs/cond-mat/0509786}{{\ttfamily
  cond-mat/0509786}}].

\bibitem{Nishida:2006br}
Y.~Nishida and D.~T. Son, \emph{{An Epsilon expansion for Fermi gas at infinite
  scattering length}},
  \href{https://doi.org/10.1103/PhysRevLett.97.050403}{\emph{Phys. Rev. Lett.}
  {\bfseries 97} (2006) 050403}
  [\href{https://arxiv.org/abs/cond-mat/0604500}{{\ttfamily
  cond-mat/0604500}}].

\bibitem{Nishida:2006eu}
Y.~Nishida and D.~T. Son, \emph{{Fermi gas near unitarity around four and two
  spatial dimensions}},
  \href{https://doi.org/10.1103/PhysRevA.75.063617}{\emph{Phys. Rev.}
  {\bfseries A75} (2007) 063617}
  [\href{https://arxiv.org/abs/cond-mat/0607835}{{\ttfamily
  cond-mat/0607835}}].

\bibitem{Nikolic:2007zz}
P.~Nikolic and S.~Sachdev, \emph{{Renormalization-group fixed points, universal
  phase diagram, and 1/N expansion for quantum liquids with interactions near
  the unitarity limit}},
  \href{https://doi.org/10.1103/PhysRevA.75.033608}{\emph{Phys. Rev.}
  {\bfseries A75} (2007) 033608}
  [\href{https://arxiv.org/abs/cond-mat/0609106}{{\ttfamily
  cond-mat/0609106}}].

\bibitem{Hagen:1983rp}
C.~Hagen, \emph{{A New Gauge Theory Without an Elementary Photon}},
  \href{https://doi.org/10.1016/0003-4916(84)90064-2}{\emph{Annals Phys.}
  {\bfseries 157} (1984) 342}.

\bibitem{Hagen:1984mj}
C.~Hagen, \emph{{Galilean-invariant gauge theory}},
  \href{https://doi.org/10.1103/PhysRevD.31.848}{\emph{Phys. Rev. D} {\bfseries
  31} (1985) 848}.

\bibitem{Jackiw:1990mb}
R.~Jackiw and S.-Y. Pi, \emph{{Classical and quantal nonrelativistic
  Chern-Simons theory}}, \href{https://doi.org/10.1103/PhysRevD.42.3500,
  10.1103/PhysRevD.48.3929}{\emph{Phys. Rev.} {\bfseries D42} (1990) 3500}.

\bibitem{Bergman}
O.~Bergman and G.~Lozano, \emph{{Aharonov-Bohm scattering, contact interactions
  and scale invariance}},
  \href{https://doi.org/10.1006/aphy.1994.1013}{\emph{Annals Phys.} {\bfseries
  229} (1994) 416} [\href{https://arxiv.org/abs/hep-th/9302116}{{\ttfamily
  hep-th/9302116}}].

\bibitem{Doroud:2015fsz}
N.~Doroud, D.~Tong and C.~Turner, \emph{{On Superconformal Anyons}},
  \href{https://doi.org/10.1007/JHEP01(2016)138}{\emph{JHEP} {\bfseries 01}
  (2016) 138} [\href{https://arxiv.org/abs/1511.01491}{{\ttfamily
  1511.01491}}].

\bibitem{Doroud:2016mfv}
N.~Doroud, D.~Tong and C.~Turner, \emph{{The Conformal Spectrum of Non-Abelian
  Anyons}}, \href{https://doi.org/10.21468/SciPostPhys.4.4.022}{\emph{SciPost
  Phys.} {\bfseries 4} (2018) 022}
  [\href{https://arxiv.org/abs/1611.05848}{{\ttfamily 1611.05848}}].

\bibitem{le1973galilean}
M.~Le~Bellac and J.-M. L{\'e}vy-Leblond, \emph{Galilean electromagnetism},
  {\emph{Il Nuovo Cimento B (1971-1996)} {\bfseries 14} (1973) 217}.

\bibitem{Santos:2004pq}
E.~S. Santos, M.~de~Montigny, F.~C. Khanna and A.~E. Santana, \emph{{Galilean
  covariant Lagrangian models}},
  \href{https://doi.org/10.1088/0305-4470/37/41/011}{\emph{J. Phys.} {\bfseries
  A37} (2004) 9771}.

\bibitem{Festuccia}
G.~Festuccia, D.~Hansen, J.~Hartong and N.~A. Obers, \emph{{Symmetries and
  Couplings of Non-Relativistic Electrodynamics}},
  \href{https://doi.org/10.1007/JHEP11(2016)037}{\emph{JHEP} {\bfseries 11}
  (2016) 037} [\href{https://arxiv.org/abs/1607.01753}{{\ttfamily
  1607.01753}}].

\bibitem{Bagchi:2014ysa}
A.~Bagchi, R.~Basu and A.~Mehra, \emph{{Galilean Conformal Electrodynamics}},
  \href{https://doi.org/10.1007/JHEP11(2014)061}{\emph{JHEP} {\bfseries 11}
  (2014) 061} [\href{https://arxiv.org/abs/1408.0810}{{\ttfamily 1408.0810}}].

\bibitem{Duval:2009vt}
C.~Duval and P.~A. Horvathy, \emph{{Non-relativistic conformal symmetries and
  Newton-Cartan structures}},
  \href{https://doi.org/10.1088/1751-8113/42/46/465206}{\emph{J. Phys.}
  {\bfseries A42} (2009) 465206}
  [\href{https://arxiv.org/abs/0904.0531}{{\ttfamily 0904.0531}}].

\bibitem{Bergshoeff:2015sic}
E.~Bergshoeff, J.~Rosseel and T.~Zojer, \emph{{Non-relativistic fields from
  arbitrary contracting backgrounds}},
  \href{https://doi.org/10.1088/0264-9381/33/17/175010}{\emph{Class. Quant.
  Grav.} {\bfseries 33} (2016) 175010}
  [\href{https://arxiv.org/abs/1512.06064}{{\ttfamily 1512.06064}}].

\bibitem{ketov2013quantum}
S.~V. Ketov, \emph{Quantum non-linear sigma-models: from quantum field theory
  to supersymmetry, conformal field theory, black holes and strings}. Springer
  Science \& Business Media, 2013.

\bibitem{Leigh:1995ep}
R.~G. Leigh and M.~J. Strassler, \emph{{Exactly marginal operators and duality
  in four-dimensional N=1 supersymmetric gauge theory}},
  \href{https://doi.org/10.1016/0550-3213(95)00261-P}{\emph{Nucl. Phys.}
  {\bfseries B447} (1995) 95}
  [\href{https://arxiv.org/abs/hep-th/9503121}{{\ttfamily hep-th/9503121}}].

\bibitem{Green:2010da}
D.~Green, Z.~Komargodski, N.~Seiberg, Y.~Tachikawa and B.~Wecht, \emph{{Exactly
  Marginal Deformations and Global Symmetries}},
  \href{https://doi.org/10.1007/JHEP06(2010)106}{\emph{JHEP} {\bfseries 06}
  (2010) 106} [\href{https://arxiv.org/abs/1005.3546}{{\ttfamily 1005.3546}}].

\bibitem{Bashmakov:2017rko}
V.~Bashmakov, M.~Bertolini and H.~Raj, \emph{{On non-supersymmetric conformal
  manifolds: field theory and holography}},
  \href{https://doi.org/10.1007/JHEP11(2017)167}{\emph{JHEP} {\bfseries 11}
  (2017) 167} [\href{https://arxiv.org/abs/1709.01749}{{\ttfamily
  1709.01749}}].

\bibitem{Behan:2017mwi}
C.~Behan, \emph{{Conformal manifolds: ODEs from OPEs}},
  \href{https://doi.org/10.1007/JHEP03(2018)127}{\emph{JHEP} {\bfseries 03}
  (2018) 127} [\href{https://arxiv.org/abs/1709.03967}{{\ttfamily
  1709.03967}}].

\bibitem{Hollands:2017chb}
S.~Hollands, \emph{{Action principle for OPE}},
  \href{https://doi.org/10.1016/j.nuclphysb.2017.11.013}{\emph{Nucl. Phys.}
  {\bfseries B926} (2018) 614}
  [\href{https://arxiv.org/abs/1710.05601}{{\ttfamily 1710.05601}}].

\bibitem{Sen:2017gfr}
K.~Sen and Y.~Tachikawa, \emph{{First-order conformal perturbation theory by
  marginal operators}},  \href{https://arxiv.org/abs/1711.05947}{{\ttfamily
  1711.05947}}.

\bibitem{Gurdogan:2015csr}
{\" O}.~G{\"u}rdo{\u g}an and V.~Kazakov, \emph{{New Integrable 4D Quantum
  Field Theories from Strongly Deformed Planar $\mathcal N = $ 4 Supersymmetric
  Yang-Mills Theory}}, \href{https://doi.org/10.1103/PhysRevLett.117.201602,
  10.1103/PhysRevLett.117.259903}{\emph{Phys. Rev. Lett.} {\bfseries 117}
  (2016) 201602} [\href{https://arxiv.org/abs/1512.06704}{{\ttfamily
  1512.06704}}].

\bibitem{Grabner:2017pgm}
D.~Grabner, N.~Gromov, V.~Kazakov and G.~Korchemsky, \emph{{Strongly
  $\gamma$-Deformed $\mathcal{N}=4$ Supersymmetric Yang-Mills Theory as an
  Integrable Conformal Field Theory}},
  \href{https://doi.org/10.1103/PhysRevLett.120.111601}{\emph{Phys. Rev. Lett.}
  {\bfseries 120} (2018) 111601}
  [\href{https://arxiv.org/abs/1711.04786}{{\ttfamily 1711.04786}}].

\bibitem{Herzog:2017xha}
C.~P. Herzog and K.-W. Huang, \emph{{Boundary Conformal Field Theory and a
  Boundary Central Charge}},
  \href{https://doi.org/10.1007/JHEP10(2017)189}{\emph{JHEP} {\bfseries 10}
  (2017) 189} [\href{https://arxiv.org/abs/1707.06224}{{\ttfamily
  1707.06224}}].

\bibitem{DiPietro:2019hqe}
L.~Di~Pietro, D.~Gaiotto, E.~Lauria and J.~Wu, \emph{{3d Abelian Gauge Theories
  at the Boundary}}, \href{https://doi.org/10.1007/JHEP05(2019)091}{\emph{JHEP}
  {\bfseries 05} (2019) 091}
  [\href{https://arxiv.org/abs/1902.09567}{{\ttfamily 1902.09567}}].

\bibitem{Herzog:2019bom}
C.~P. Herzog and I.~Shamir, \emph{{On Marginal Operators in Boundary Conformal
  Field Theory}}, \href{https://doi.org/10.1007/JHEP10(2019)088}{\emph{JHEP}
  {\bfseries 10} (2019) 088}
  [\href{https://arxiv.org/abs/1906.11281}{{\ttfamily 1906.11281}}].

\bibitem{Chai:2020zgq}
N.~Chai, S.~Chaudhuri, C.~Choi, Z.~Komargodski, E.~Rabinovici and M.~Smolkin,
  \emph{{Thermal Order in Conformal Theories}},
  \href{https://arxiv.org/abs/2005.03676}{{\ttfamily 2005.03676}}.

\bibitem{Arav:2019tqm}
I.~Arav, Y.~Oz and A.~Raviv-Moshe, \emph{{Holomorphic Structure and Quantum
  Critical Points in Supersymmetric Lifshitz Field Theories}},
  \href{https://doi.org/10.1007/JHEP11(2019)064}{\emph{JHEP} {\bfseries 11}
  (2019) 064} [\href{https://arxiv.org/abs/1908.03220}{{\ttfamily
  1908.03220}}].

\bibitem{Jensen:2014aia}
K.~Jensen, \emph{{On the coupling of Galilean-invariant field theories to
  curved spacetime}},
  \href{https://doi.org/10.21468/SciPostPhys.5.1.011}{\emph{SciPost Phys.}
  {\bfseries 5} (2018) 011} [\href{https://arxiv.org/abs/1408.6855}{{\ttfamily
  1408.6855}}].

\bibitem{Son:2008ye}
D.~T. Son, \emph{{Toward an AdS/cold atoms correspondence: A Geometric
  realization of the Schrodinger symmetry}},
  \href{https://doi.org/10.1103/PhysRevD.78.046003}{\emph{Phys. Rev.}
  {\bfseries D78} (2008) 046003}
  [\href{https://arxiv.org/abs/0804.3972}{{\ttfamily 0804.3972}}].

\bibitem{Nakayama:2009ww}
Y.~Nakayama, \emph{{Gravity Dual for Reggeon Field Theory and Non-linear
  Quantum Finance}},
  \href{https://doi.org/10.1142/S0217751X09047594}{\emph{Int. J. Mod. Phys.}
  {\bfseries A24} (2009) 6197}
  [\href{https://arxiv.org/abs/0906.4112}{{\ttfamily 0906.4112}}].

\bibitem{Nakayama:2013is}
Y.~Nakayama, \emph{{Scale invariance vs conformal invariance}},
  \href{https://doi.org/10.1016/j.physrep.2014.12.003}{\emph{Phys. Rept.}
  {\bfseries 569} (2015) 1} [\href{https://arxiv.org/abs/1302.0884}{{\ttfamily
  1302.0884}}].

\bibitem{Arav:2016xjc}
I.~Arav, S.~Chapman and Y.~Oz, \emph{{Non-Relativistic Scale Anomalies}},
  \href{https://doi.org/10.1007/JHEP06(2016)158}{\emph{JHEP} {\bfseries 06}
  (2016) 158} [\href{https://arxiv.org/abs/1601.06795}{{\ttfamily
  1601.06795}}].

\bibitem{Jensen:2014wha}
K.~Jensen and A.~Karch, \emph{{Revisiting non-relativistic limits}},
  \href{https://doi.org/10.1007/JHEP04(2015)155}{\emph{JHEP} {\bfseries 04}
  (2015) 155} [\href{https://arxiv.org/abs/1412.2738}{{\ttfamily 1412.2738}}].

\bibitem{Banerjee:2019axy}
K.~Banerjee, R.~Basu and A.~Mohan, \emph{{Uniqueness of Galilean Conformal
  Electrodynamics and its Dynamical Structure}},
  \href{https://arxiv.org/abs/1909.11993}{{\ttfamily 1909.11993}}.

\bibitem{Manohar:2018aog}
A.~V. Manohar, \emph{{Introduction to Effective Field Theories}},  in
  \emph{{Les Houches summer school: EFT in Particle Physics and Cosmology Les
  Houches, Chamonix Valley, France, July 3-28, 2017}}, 2018,
  \href{https://arxiv.org/abs/1804.05863}{{\ttfamily 1804.05863}}.

\bibitem{Bergman:1991hf}
O.~Bergman, \emph{{Nonrelativistic field theoretic scale anomaly}},
  \href{https://doi.org/10.1103/PhysRevD.46.5474}{\emph{Phys. Rev. D}
  {\bfseries 46} (1992) 5474}.

\bibitem{Klose:2006dd}
T.~Klose and K.~Zarembo, \emph{{Bethe ansatz in stringy sigma models}},
  \href{https://doi.org/10.1088/1742-5468/2006/05/P05006}{\emph{J. Stat. Mech.}
  {\bfseries 0605} (2006) P05006}
  [\href{https://arxiv.org/abs/hep-th/0603039}{{\ttfamily hep-th/0603039}}].

\bibitem{Auzzi:2019kdd}
R.~Auzzi, S.~Baiguera, G.~Nardelli and S.~Penati, \emph{{Renormalization
  properties of a Galilean Wess-Zumino model}},
  \href{https://doi.org/10.1007/JHEP06(2019)048}{\emph{JHEP} {\bfseries 06}
  (2019) 048} [\href{https://arxiv.org/abs/1904.08404}{{\ttfamily
  1904.08404}}].

\bibitem{Caswell:1985ui}
W.~Caswell and G.~Lepage, \emph{{Effective Lagrangians for Bound State Problems
  in QED, QCD, and Other Field Theories}},
  \href{https://doi.org/10.1016/0370-2693(86)91297-9}{\emph{Phys. Lett. B}
  {\bfseries 167} (1986) 437}.

\bibitem{Labelle:1996en}
P.~Labelle, \emph{{Effective field theories for QED bound states: Extending
  nonrelativistic QED to study retardation effects}},
  \href{https://doi.org/10.1103/PhysRevD.58.093013}{\emph{Phys. Rev. D}
  {\bfseries 58} (1998) 093013}
  [\href{https://arxiv.org/abs/hep-ph/9608491}{{\ttfamily hep-ph/9608491}}].

\bibitem{Leibbrandt:1996np}
G.~Leibbrandt and J.~Williams, \emph{{Split dimensional regularization for the
  Coulomb gauge}},
  \href{https://doi.org/10.1016/0550-3213(96)00299-4}{\emph{Nucl. Phys. B}
  {\bfseries 475} (1996) 469}
  [\href{https://arxiv.org/abs/hep-th/9601046}{{\ttfamily hep-th/9601046}}].

\bibitem{Leibbrandt:1997kh}
G.~Leibbrandt, \emph{{The three point function in split dimensional
  regularization in the Coulomb gauge}},
  \href{https://doi.org/10.1016/S0550-3213(98)00211-9}{\emph{Nucl. Phys. B}
  {\bfseries 521} (1998) 383}
  [\href{https://arxiv.org/abs/hep-th/9804109}{{\ttfamily hep-th/9804109}}].

\bibitem{Anselmi:2007ri}
D.~Anselmi and M.~Halat, \emph{{Renormalization of Lorentz violating
  theories}}, \href{https://doi.org/10.1103/PhysRevD.76.125011}{\emph{Phys.
  Rev. D} {\bfseries 76} (2007) 125011}
  [\href{https://arxiv.org/abs/0707.2480}{{\ttfamily 0707.2480}}].

\bibitem{Arav:2016akx}
I.~Arav, Y.~Oz and A.~Raviv-Moshe, \emph{{Lifshitz Anomalies, Ward Identities
  and Split Dimensional Regularization}},
  \href{https://doi.org/10.1007/JHEP03(2017)088}{\emph{JHEP} {\bfseries 03}
  (2017) 088} [\href{https://arxiv.org/abs/1612.03500}{{\ttfamily
  1612.03500}}].

\bibitem{Bagchi:2015qcw}
A.~Bagchi, R.~Basu, A.~Kakkar and A.~Mehra, \emph{{Galilean Yang-Mills
  Theory}}, \href{https://doi.org/10.1007/JHEP04(2016)051}{\emph{JHEP}
  {\bfseries 04} (2016) 051}
  [\href{https://arxiv.org/abs/1512.08375}{{\ttfamily 1512.08375}}].

\bibitem{Gomis:2020fui}
J.~Gomis, Z.~Yan and M.~Yu, \emph{{Nonrelativistic Open String and Yang-Mills
  Theory}},  \href{https://arxiv.org/abs/2007.01886}{{\ttfamily 2007.01886}}.

\bibitem{SUSYGED}
S.~Chapman, Y.~Oz and A.~Raviv-Moshe, \emph{{Supersymetric Galilean
  Electrodynamics}},  in progress.

\bibitem{Gomis:2000bd}
J.~Gomis and H.~Ooguri, \emph{{Nonrelativistic closed string theory}},
  \href{https://doi.org/10.1063/1.1372697}{\emph{J. Math. Phys.} {\bfseries 42}
  (2001) 3127} [\href{https://arxiv.org/abs/hep-th/0009181}{{\ttfamily
  hep-th/0009181}}].

\bibitem{Danielsson:2000gi}
U.~H. Danielsson, A.~Guijosa and M.~Kruczenski, \emph{{IIA/B, wound and
  wrapped}}, \href{https://doi.org/10.1088/1126-6708/2000/10/020}{\emph{JHEP}
  {\bfseries 10} (2000) 020}
  [\href{https://arxiv.org/abs/hep-th/0009182}{{\ttfamily hep-th/0009182}}].

\bibitem{Danielsson:2000mu}
U.~H. Danielsson, A.~Guijosa and M.~Kruczenski, \emph{{Newtonian gravitons and
  d-brane collective coordinates in wound string theory}},
  \href{https://doi.org/10.1088/1126-6708/2001/03/041}{\emph{JHEP} {\bfseries
  03} (2001) 041} [\href{https://arxiv.org/abs/hep-th/0012183}{{\ttfamily
  hep-th/0012183}}].

\bibitem{Bergshoeff:2019pij}
E.~A. Bergshoeff, J.~Gomis, J.~Rosseel, C.~\c{S}im\c{s}ek and Z.~Yan,
  \emph{{String Theory and String Newton-Cartan Geometry}},
  \href{https://doi.org/10.1088/1751-8121/ab56e9}{\emph{J. Phys.} {\bfseries
  A53} (2020) 014001} [\href{https://arxiv.org/abs/1907.10668}{{\ttfamily
  1907.10668}}].

\bibitem{Berthier:2017slt}
L.~Berthier, K.~T. Grosvenor and Z.~Yan, \emph{{Nonrelativistic Yang-Mills
  Theory for a Naturally Light Higgs Boson}},
  \href{https://doi.org/10.1103/PhysRevD.96.095030}{\emph{Phys. Rev.}
  {\bfseries D96} (2017) 095030}
  [\href{https://arxiv.org/abs/1705.04701}{{\ttfamily 1705.04701}}].

\end{thebibliography}\endgroup
